\documentclass{aa}  

\usepackage{graphicx}
\usepackage{txfonts}
\usepackage{hyperref}
\usepackage{url}

\def\mobs{M_*^{\mathrm{(obs)}}}

\def\mstar{M_*}

\def\mten{M_{*,10}}
\def\mtenobs{M_{*,10}^{\mathrm{(obs)}}}
\def\mtenobsmin{M_{\mathrm{min}}}

\def\slope{\Gamma_{*,10}}
\def\slopeobs{\Gamma_{*,10}^{\mathrm{(obs)}}}

\def\reff{R_{\mathrm{e}}}

\def\rap{R_{\mathrm{ap}}}

\def\rsky{R_{\mathrm{sky}}}
\def\fdet{f_{\mathrm{det}}}

\def\nser{n}

\def\sigmaetwo{\sigma_{\mathrm{e}2}}
\def\sigmaap{\sigma_{\mathrm{ap}}}
\def\sigmaten{\sigma_{10}}

\def\mrpetro{m_{r,p}^{\mathrm{(SDSS)}}}
\def\mrten{m_{r,10}^{\mathrm{(HSC)}}}

\def\Sref#1{Section~\ref{#1}\xspace}
\def\Fref#1{Figure~\ref{#1}\xspace}
\def\Tref#1{Table~\ref{#1}\xspace}
\def\Eref#1{Equation~\ref{#1}\xspace}

\def\Nsdss{2,125}
\def\Ngama{776}

\def\eagle{{\sc EAGLE}}

\begin{document}

   \title{A robust two-parameter description of the stellar profile of elliptical galaxies.}
   \titlerunning{A robust description of elliptical galaxies.}
   \authorrunning{Sonnenfeld}

   %\subtitle{I. Overviewing the $\kappa$-mechanism}

   \author{Alessandro Sonnenfeld\inst{1}\thanks{Marie Sk\l{}odowska-Curie Fellow}
          }

   \institute{Leiden Observatory, Leiden University, Niels Bohrweg 2, 2333 CA Leiden, the Netherlands\\
              \email{sonnenfeld@strw.leidenuniv.nl}
             }

   \date{}

% \abstract{}{}{}{}{} 
% 5 {} token are mandatory
 
  \abstract
  % context heading (optional)
  % {} leave it empty if necessary  
   %{Massive elliptical galaxies are not homologous systems: at fixed stellar mass, variations in the half-light radius translate in changes in their stellar velocity dispersion that are weaker than suggested by a simple virial scaling relation, a feature generally known as the tilt of the fundamental plane. This suggests that at least two parameters are necessary to describe this population of objects.
   %{Massive elliptical galaxies are not homologous systems: at least two parameters are necessary to describe this population of objects. Generally, stellar mass and half-light radius are used for this purpose. The total mass of a galaxy, however, is not a well-defined quantity, due to the finite depth of photometric observations and the arbitrariness of the distinction between galaxy and diffuse intra-group light. This limits our ability to make accurate comparisons between models and observations.}
   {The stellar density profile a galaxy is typically summarised with two numbers: total stellar mass and half-light radius. The total mass of a galaxy, however, is not a well-defined quantity, due to the finite depth of photometric observations and the arbitrariness of the distinction between galaxy and diffuse intra-group light. This limits our ability to make accurate comparisons between models and observations.}
  % aims heading (mandatory)
   {
    I wish to provide a more robust two-parameter description of the stellar density distribution of elliptical galaxies, in terms of quantities that can be measured unambiguously.
} 
   % methods heading (mandatory)
   {
I propose to use the stellar mass enclosed within 10~kpc in projection, $\mten$, and the mass-weighted stellar density slope within the same aperture, $\slope$, for this purpose.
I measured the distribution in $\mten$ and $\slope$ of a sample of elliptical galaxies from the Sloan Digital Sky Survey and the Galaxy And Mass Assembly survey, using photometry from the Hyper Suprime-Cam survey.
I measured, at fixed $(\mten,\slope)$, what is the spread in galaxy surface brightness profile and central stellar velocity dispersion within the sample.
As a first application, I then compared the observed $\mten-\slope$ relation of elliptical galaxies with that of similarly selected galaxies in the \eagle\ {\sc Reference} simulation.
}
% results heading (mandatory)
   {
The pair of values of $(\mten,\slope)$ can be used to predict the stellar density profile in the inner 10~kpc of a galaxy with better than 20\% accuracy. Similarly, $\mten$ and $\slope$ can be combined to obtain a proxy for stellar velocity dispersion at least as good as the stellar mass fundamental plane.
The average stellar density slope of \eagle\ elliptical galaxies matches well that of observed ones at $\mten=10^{11}M_\odot$, but the \eagle\ $\mten-\slope$ relation is shallower and has a larger intrinsic scatter compared to observations.
}
  % conclusions heading (optional), leave it empty if necessary 
   {
This new parameterisation of the stellar density profile of massive elliptical galaxies provides a more robust way of comparing results from different photometric surveys and from hydrodynamical simulations, with respect to a description based on total stellar mass and half-light radius.
}

   \maketitle
%
%________________________________________________________________

\section{Introduction}\label{sect:intro}

%Massive elliptical galaxies are well-behaved objects, but their evolution is still not fully understood.
%One way to make progress is to compare observations with hydro simulations.
%However, systematics in the way the stellar distribution of galaxies are measured in observations are limiting our ability to make precise comparisons.

Elliptical galaxies have, by definition, a smooth surface brightness distribution,
the shape of which is determined by their assembly history \citep{NJO09, Hop++10, HNO13}. For this reason, there have been numerous studies aimed at understanding the evolution of elliptical galaxies by means of their surface brightness distribution and its variation across the galaxy population \citep[e.g.][]{vDo++10, Ber++11, New++12, Nip++12, Hue++13, Sha++14, Fag++16, YKJ17, SWB19, Hua++20}.
%The shape of their surface brightness profile is determined by their assembly history \citep{Hil++11} and therefore holds important clues regarding what physical processes affect their evolution.
Traditionally, the surface brightness profile of a galaxy is summarised by two numbers: the total integrated light (or mass, if an estimate of the stellar mass-to-light ratio is available) and the half-light radius.
However, both of these quantities suffer from extrapolation problems: the finite depth of photometric observations limits our ability to measure the light of a galaxy to a region where its surface brightness is above the background noise level. As a result, the total luminosity needs to be defined either by extrapolating the surface brightness profile to regions not constrained by the data, or by imposing an arbitrary truncation radius.
%The problem becomes increasingly more important for higher redshift galaxies, but as I'll show, it's already non-negligible at $z=0.2$.
Ultimately, the total luminosity of a galaxy, and consequently the half-light radius, is not a well-defined quantity because of the arbitrary distinction between stars belonging to it and those associated with the intra-group medium.
This ambiguity in the definition of the stellar mass is a problem, because it limits our ability to make meaningful comparisons between theoretical models and observations.
One such example is the recent investigation of the correlation between the surface brightness profiles of massive galaxies and the mass of their dark matter halo, at fixed stellar mass: \citet{SWB19} and \citet{Hua++20} used essentially the same data but different definitions of the stellar mass of a galaxy, reaching different conclusions.

In this work, I propose an alternative approach to the problem of characterising the stellar distribution of massive elliptical galaxies and related properties, in which I give up attempting to capture the total mass of a galaxy, and instead focus on a region well constrained by the data.
This new description is based on two parameters: the projected stellar mass enclosed within a circularised aperture of radius $10$~kpc, $\mten$, and the stellar mass-weighted projected slope measured within the same aperture, $\slope$. The latter is defined as
\begin{equation}\label{eq:slope}
\slope \equiv - \dfrac{2\pi\int_0^{10}R\dfrac{d\log \Sigma_*}{d\log R}\Sigma_*(R)dR}{2\pi\int_0^{10}R\Sigma_*(R)dR} = 2 - \dfrac{2\pi(10)^2\Sigma_*(10)}{\mten},
\end{equation}
where $\Sigma_*(R)$ is the stellar surface mass density profile and radii are expressed in units of kpc.
The choice of 10~kpc as the reference scale is a compromise between the will to select an aperture that is large enough to enclose a significant fraction of the total stellar mass of a galaxy (however this is defined) but not too large that its surface brightness is below the background noise level at that radius.

I measure the values of $(\mten,\slope)$ space of a sample of massive quiescent galaxies with elliptical morphology, selected from the Sloan Digital Sky Survey \citep[SDSS]{Yor++00} spectroscopic sample and the Gama And Mass Assembly survey \citep[GAMA][]{Dri++09,Bal++10,Rob++10}, using photometric data from the Hyper Suprime-Cam \citep[HSC]{Miy++18,Kom++18,Kaw++18} Subaru Strategic Program \citep[][HSC survey from here on]{Aih++18}.
I focus on two questions: when fixing $\mten$ and $\slope$, how much residual variation is there in the surface brightness profile of massive elliptical galaxies? How well do $\mten$ and $\slope$ predict the stellar velocity dispersion of these galaxies, compared to the total stellar mass and the half-light radius (i.e. the stellar mass fundamental plane)?
I then show, as an example of an application of this new parameterisation of the stellar density profile, a comparison between the observed $\mten-\slope$ distribution with that of simulated galaxies.

The structure of this paper is the following. In \Sref{sect:data} I describe the data used for this study, including sample selection criteria, and present measurements of the stellar density profile of the galaxies in the sample. In \Sref{sect:m10slope}, I measure values of $\mten$ and $\slope$ and investigate how well they can predict the surface brightness profile of massive galaxies, and out to what scales. In \Sref{sect:veldisp}, I infer the distribution of stellar velocity dispersion of the galaxy in the sample as a function of stellar mass and size on the one hand, and $\mten$ and $\slope$ on the other hand. 
In \Sref{sect:eagle}, I select a volume-limited subsample of galaxies and compare their $\mten-\slope$ relation with that of galaxies from the \eagle\ hydrodynamical simulation \citep{Sch++15}.
I discuss the results in \Sref{sect:discuss} and summarise in \Sref{sect:concl}.

I assume a flat $\Lambda$CDM cosmology with $\Omega_M=0.3$ and $H_0=70\,\rm{km}\,\rm{s}^{-1}\,\rm{Mpc}^{-1}$. Magnitudes are in AB units, stellar masses are in solar units and velocity dispersions are in $\rm{km}\,\rm{s}^{-1}$.
%I make the code and Markov Chain Monte Carlo (MCMC) samples of the fits of population models available online\footnote{\url{https://github.com/astrosonnen/m10_slope/}}.

%__________________________________________________________________

\section{Data}\label{sect:data}

\subsection{Sample selection}\label{ssec:sample}

For this study, I focus on quiescent galaxies with elliptical morphology.
I consider two samples of galaxies: the first one drawn from the SDSS spectroscopic sample and the second one from the GAMA survey. 
After showing measurements of $\mten$ and $\slope$ carried out on both samples, I will use the SDSS sample to measure the distribution in stellar velocity dispersion as a function of $\mten$ and $\slope$, and the GAMA sample to infer the distribution of $\slope$ as a function of $\mten$ on a complete subset of objects.
%I will show measurements of $\mten$ and $\slope$ carried out on galaxies from both samples. Then, I will use the SDSS sample to measure the distribution in stellar velocity dispersion as a function of $\mten$ and $\slope$.
%The first one is drawn from the SDSS spectroscopic sample and will be used to measure the distribution in stellar velocity dispersion as a function of $\mten$ and $\slope$. 
%The second sample is drawn from the GAMA spectroscopic survey. Measurements of the stellar velocity dispersion of 

\subsubsection{The SDSS sample}

\begin{figure*}
\includegraphics[width=\textwidth]{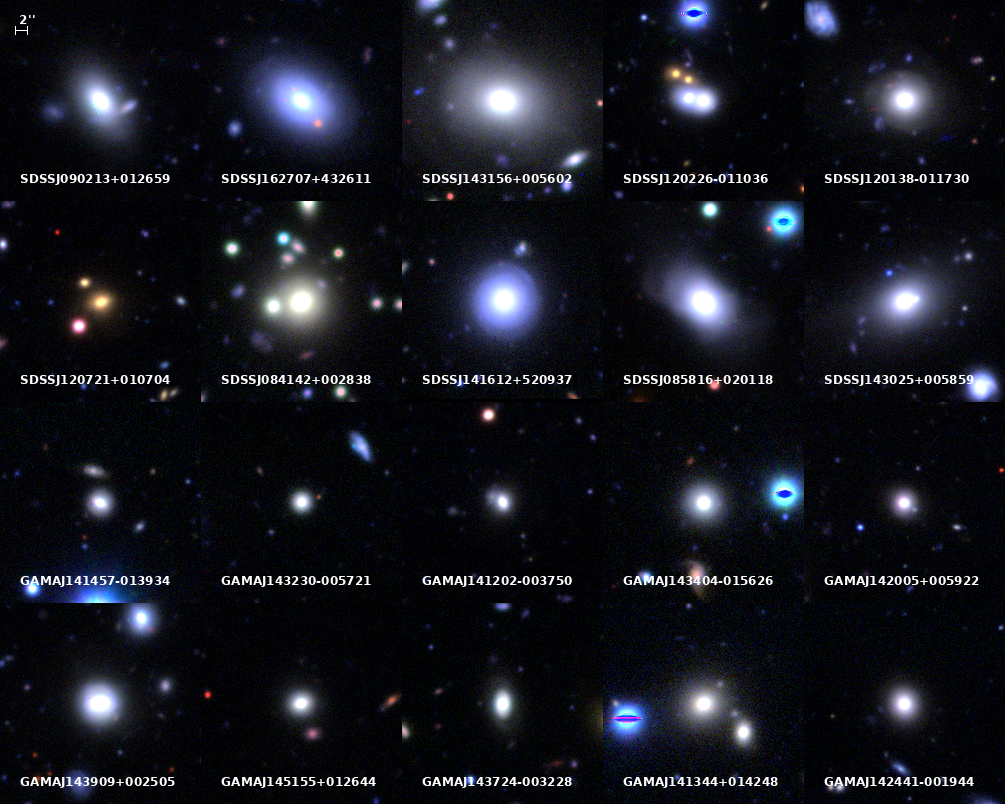}
\caption{Color-composite images, made with HSC $g-$, $r-$ and $i-$band images, of twenty randomly selected elliptical galaxies from the SDSS and GAMA samples.
\label{fig:sample}
}
\end{figure*}

I selected galaxies with spectra from the SDSS Legacy Survey\footnote{\url{sdss2.org/legacy/}}, using the spectroscopic catalog from the twelfth data release of the SDSS \citep[DR12][]{Ala++15}. The SDSS Legacy Survey includes galaxies from the Main Galaxy Sample \citep{Str++02} and then the Luminous Red Galaxy sample \citep{Eis++01} and covers the redshift range $0 < z < 0.5$.
I applied a cut on H$\alpha$ equivalent width, $\rm{EW}_{{\rm H}\alpha} > -3\AA$ (negative equivalent width corresponding to emission), to remove highly star-forming galaxies. 
I then focused on $\sim3,800$ objects with photometric data from the Wide component of the first public data release of the HSC survey \citep{Aih++18b}.
From these, I removed 783 objects flagged as saturated in at least one filter of HSC.
I then visually inspected the remaining sample and selected only galaxies with elliptical morphology and undisturbed photometry. In particular, I removed any objects with signs of spiral arms or disks, as well as objects with close neighbours or with significant contamination from a nearby bright star.
This visual inspection step reduced the sample size to $\Nsdss$.
Color-composite images of a randomly selected subset of ten SDSS galaxies are shown in the top half of \Fref{fig:sample}.

\subsubsection{The GAMA sample}

I selected galaxies from the second data release of the GAMA survey \citep[GAMA DR2,][]{Lis++15}, located in the G15 field of GAMA, with a spectroscopic redshift in the range $0.15 < z < 0.2$, an H$\alpha$ equivalent width larger than $-3\AA$ and with available photometric data from the second public data release of the HSC survey \citep[HSC PDR2][]{Aih++19}.
The lower redshift limit is applied to minimize the number of objects with saturated images (less than 5\% above this redshift), while the upper limit will be needed to define a volume-limited sample of galaxies.
Among the $1,378$ objects that satisfy the requirements listed above, I selected $\Ngama$ object visually classified as ellipticals using the same criteria adopted for the SDSS sample. 
Color-composite images of a randomly selected subset of ten GAMA galaxies are shown in the bottom half of \Fref{fig:sample}.

The main difference between the two samples is the availability of stellar velocity dispersion measurements, limited to the SDSS sample, and the depth of the parent surveys from which they are drawn. The GAMA DR2 is complete down to an $r-$band magnitude of $19.4$ in the G15 field, while the SDSS Legacy Survey is shallower.
Although the third data release of GAMA extends the G15 sample down to $r < 19.8$ \citep{Bal++18}, I only considered objects included in DR2 because they have already been cross-matched to the HSC catalog as part of the HSC PDR2.
A subset of 44 galaxies is common to both the SDSS and GAMA samples.

\subsection{HSC photometry}

In order to measure the surface brightness and stellar mass profile of the galaxies in the two samples, I used photometric data from the HSC PDR2.
For each galaxy, I obtained $201\times201$ pixel ($33.77''\times33.77''$) cutouts in $g, r, i, z, y$ bands. Each cutout consisted of coadded sky-subtracted science, variance and mask maps, produced by the HSC data reduction pipeline \citep[{\sc HSCpipe},][]{Bos++18}.

\subsection{Surface brightness profile measurements}\label{ssec:sb}

I modeled the surface brightness distribution of each galaxy with an elliptical S\'{e}rsic profile,
\begin{equation}\label{eq:sersic}
I(x,y) = I_0\exp{\left\{-b(n)\left(\frac{R}{\reff}\right)^{1/n}\right\}},
\end{equation}
where $x$ and $y$ are Cartesian coordinates with origin at the galaxy centre and aligned with the galaxy major and minor axes, $R$ is the circularised radius
\begin{equation}\label{eq:ellcoord}
R^2 \equiv qx^2 + \frac{y^2}{q},
\end{equation}
$q$ is the axis ratio, $n$ is the S\'{e}rsic index, and $b(n)$ is a numerical constant defined in such a way that the isophote of radius $R=\reff$ encloses half of the total light \citep[see][]{C+B99}.

I fitted seeing-convolved S\'{e}rsic models to sky-subtracted and coadded images in $g, r, i, z, y$ bands, following the same procedure adopted by \citet{SWB19} for the analysis of data from the 17A internal data release of the HSC survey.
The structural parameters of the S\'{e}rsic model were assumed to be the same in all bands, thereby asserting spatially constant colors.
Neighbouring galaxies were identified and masked out by running the software {\sc SExtractor} \citep{B+A96} on the $i-$band image with a $2\sigma$ object detection threshold.
Saturated pixels were masked out using the masks provided by the HSC PDR2.

As I argued in the Introduction, when fitting an analytical profile to the surface brightness distribution of a galaxy, part of the inferred total brightness is due to an extrapolation of the model to large radii not constrained by the data. 
In order to gauge the importance of this effect for the galaxies in the two samples, I measured for each object the radius $\rsky$ at which the model surface brightness equals the sky background surface brightness 1$\sigma$ fluctuation on a single pixel. I then computed the fraction $\fdet$ of the total brightness inferred from the S\'{e}rsic profile fit accounted for by the region $R<\rsky$: in other words, $\fdet$ is the fraction of the total brightness that is directly constrained by the data.
Both $\rsky$ and $\fdet$ are functions of the $i-$band magnitude, half-light radius, S\'{e}rsic index and the depth of the data.

\begin{figure*}
\includegraphics[width=\textwidth]{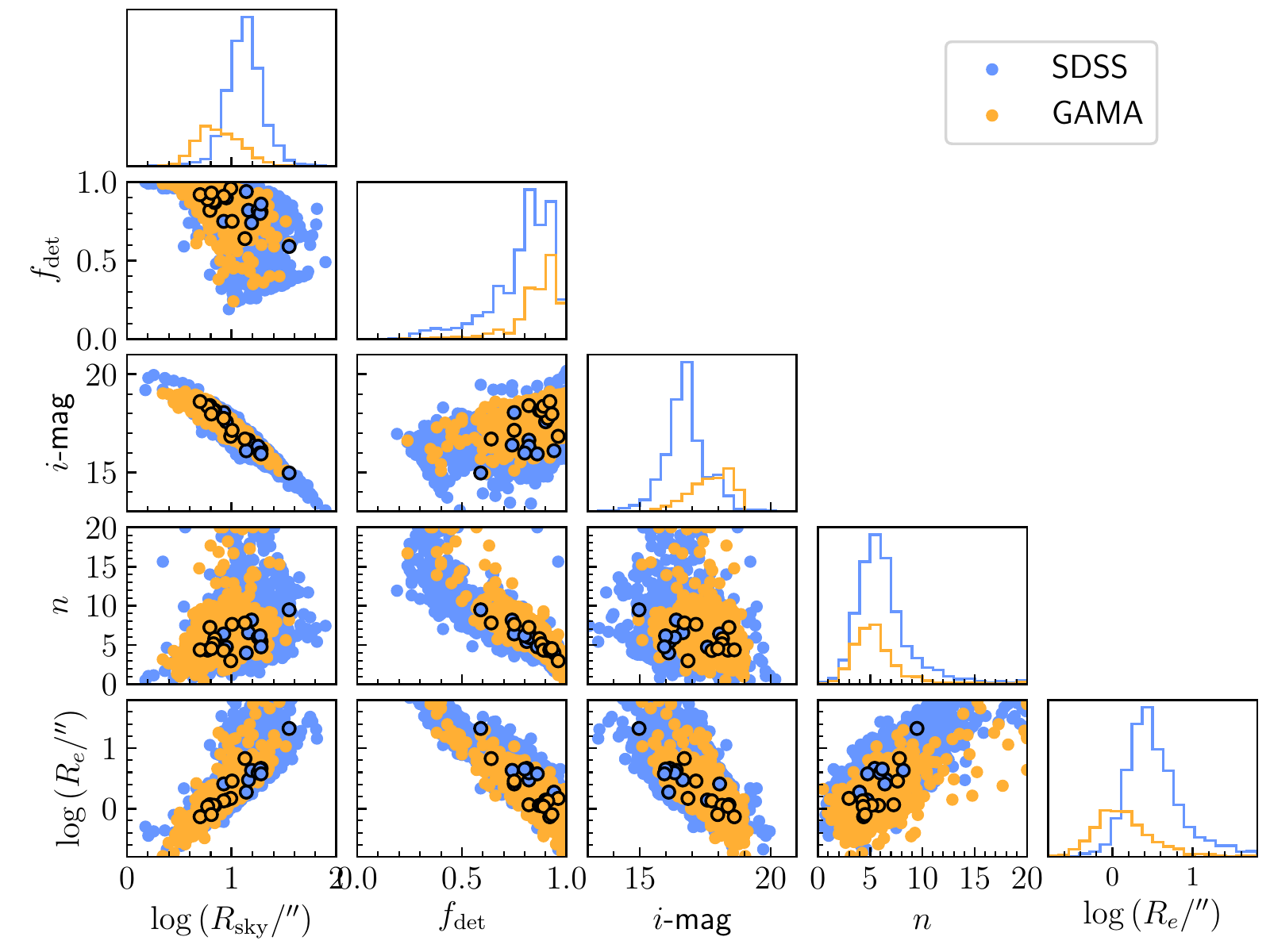}
\caption{Distribution in half-light radius, $i-$band magnitude, S\'{e}rsic index, radius $\rsky$ at which the $i-$band surface brightness of a galaxy falls below the sky fluctuation level and fraction $\fdet$ of the total $i-$band flux from S\'{e}rsic profile fitting enclosed within the isophote of radius $\rsky$.
Points circled in black correspond to the twenty galaxies shown in \Fref{fig:sample}.
}
\label{fig:rskyfdetcp}
\end{figure*}

\begin{figure}
\centering
\includegraphics[width=\columnwidth]{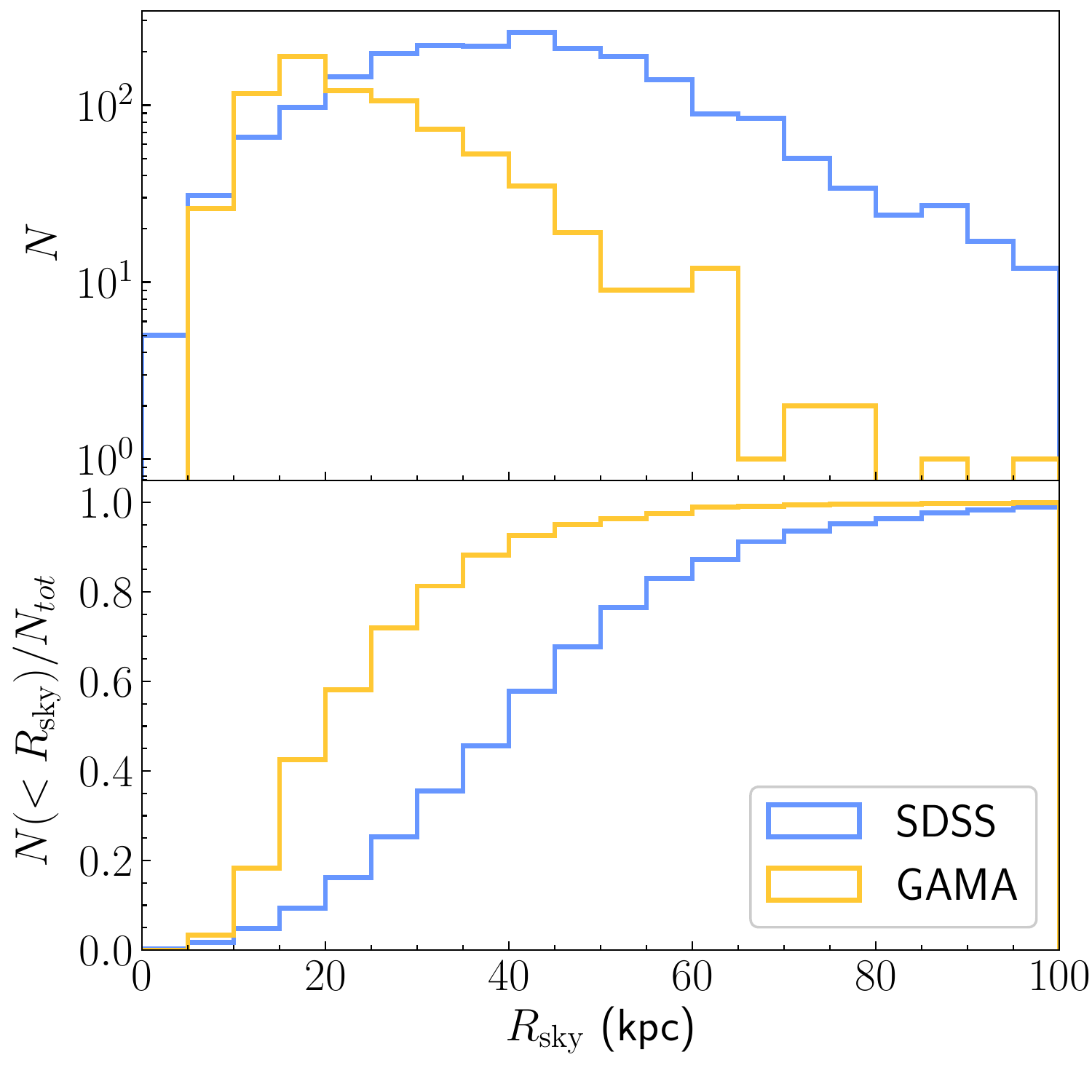}
 \caption{{\em Top:} Distribution in the radius at which the $i-$band surface brightness of a galaxy falls below the sky fluctuation level, $\rsky$, in physical units. {\em Bottom:} Cumulative distribution.}
 \label{fig:rsky_phys}
\end{figure}

The distribution of these five quantities is plotted in \Fref{fig:rskyfdetcp}. 
We can see clear anti-correlations between $\fdet$ and both S\'{e}rsic index $n$ and half-light radius $\reff$, which can be understood as follows: a larger S\'{e}rsic index corresponds to a shallower surface brightness profile as $R \rightarrow \infty$, which means that regions at large radii contribute to a larger fraction of the total light. This results in larger half-light radii (hence the positive correlation between $n$ and $\reff$) and lower values of $\fdet$.
Most importantly, for roughly a third of the SDSS sample, $\fdet$ is smaller than $0.8$, meaning that 20\% of the measured total flux or more is a result of an extrapolation of the best-fit S\'{e}rsic profile to regions not directly constrained by the data.
While this might not seem like a big systematic uncertainty, it is worth pointing out that this is a sample of relatively bright objects, with data from the currently deepest wide field photometric survey (HSC).
The typical value of $\fdet$ is lower for galaxies at a higher redshift or with shallower photometric data, compared to the distribution shown in \Fref{fig:rskyfdetcp}.
This is one of the main motivations for replacing $\mstar$ and $\reff$ with $\mten$ and $\slope$ in the description of the stellar density profile of massive elliptical galaxies.
In \Fref{fig:rsky_phys}, we show the distribution of $\rsky$ in physical units. As we can see from the top panel, all but a handful of objects from both the SDSS and GAMA samples have values of $\rsky$ smaller than $10$~kpc. This means that the measurements of $\mten$ and $\slope$ are well constrained by the data, for the galaxies in the two samples.

\subsection{Goodness of fit}\label{ssec:gof}

\begin{figure*}
\includegraphics[width=\textwidth]{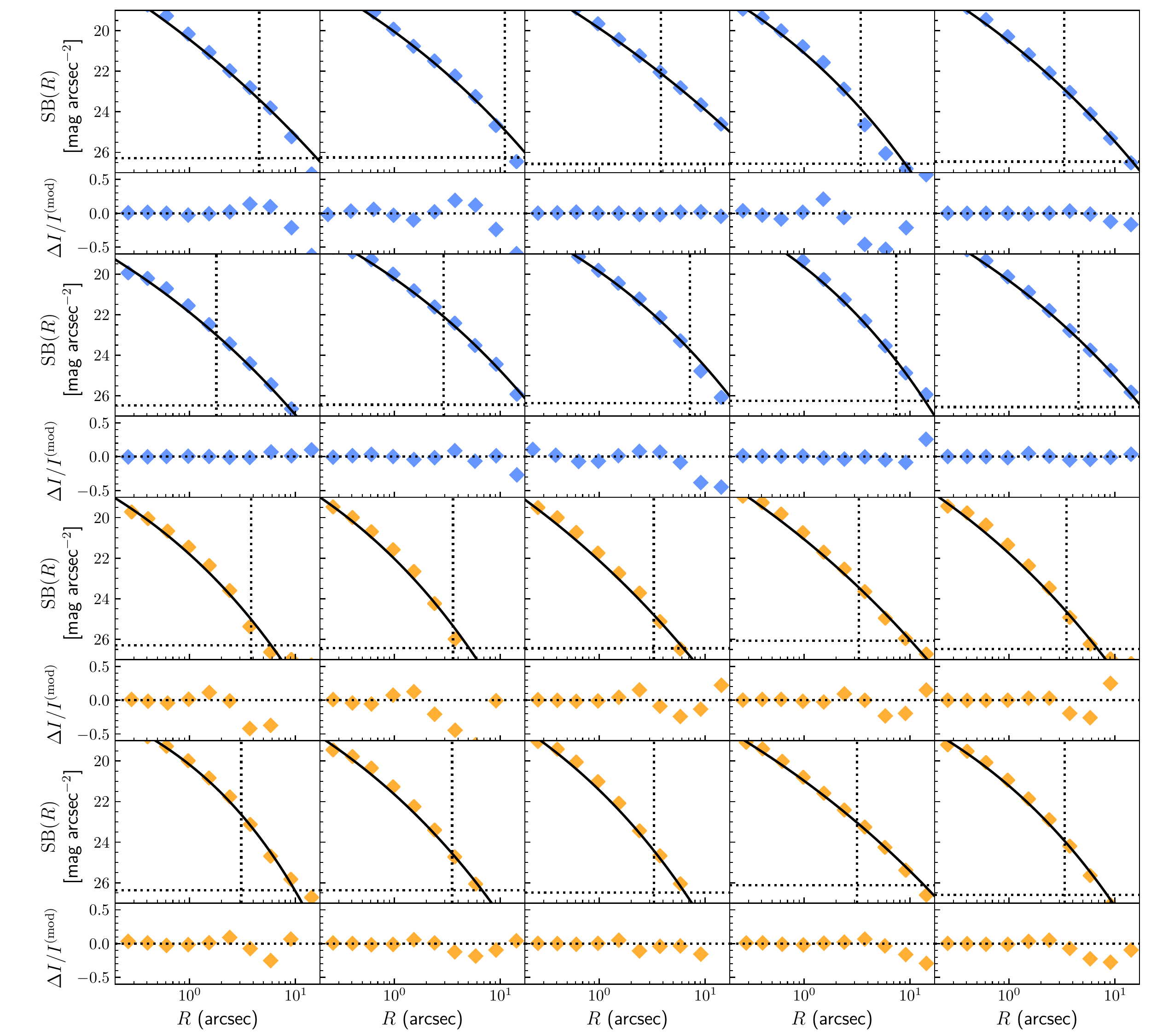}
\caption{
Surface brightness profiles in $i-$band of the twenty galaxies shown in \Fref{fig:sample}. {\em Top:} observed surface brightness, azimuthally averaged in radial bins (blue diamonds) and best-fit S\'{e}rsic model (black solid curve), as a function of circularised radius.
{\em Bottom:} residuals between observed and model flux, divided by the model flux.
The two different colours correspond to SDSS and GAMA galaxies, as in \Fref{fig:rskyfdetcp}.
%Surface brightness profiles in $i-$band of the twenty galaxies shown in \Fref{fig:sample}. {\em Top:} observed surface brightness at each unmasked pixel (blue points) and best-fit S\'{e}rsic model (black solid curve), as a function of circularised radius. The vertical dotted line corresponds to a physical distance of $10$~kpc from the galaxy center, while the horizontal dotted line marks the surface brightness level of the background noise fluctuation for a single pixel.
%{\em Bottom:} residuals between observed and model flux divided by the model flux, for each pixel (blue points) and calculated in radial bins (orange diamonds).
\label{fig:gof}
}
\end{figure*}

In the previous subsection I assumed that the surface brightness distribution of the elliptical galaxies in the two samples can be described by a S\'{e}rsic profile with uniform colours. 
In the rest of this work I will rely on this assumption to calculate the stellar mass distribution and compute $\mten$ and $\slope$. Therefore, it is important to check how well does a S\'{e}rsic model fit the observed surface brightness data.
\Fref{fig:gof} compares the observed $i-$band surface brightness profile with the best-fit S\'{e}rsic model, for the same twenty galaxies previously shown in \Fref{fig:sample}.
The S\'{e}rsic profile follows the data remarkably well: residuals between the binned data and the model (bottom panels) are almost everywhere smaller than 20\% of the model flux, with the largest deviations observed only at large radii.
%The largest deviations are observed at large radii, while in the inner 10~kpc (vertical dotted line) 
%The goodness of fit can be evaluated by looking at the bottom panel, which shows the residual between the observed data and the flux predicted by the best-fit S\'{e}rsic model, divided by the model flux itself.
%Individual pixel data are naturally very noisy and show , but when averaged azimuthally in radial bins (orange diamonds), 
%The S\'{e}rsic model can typically reproduce the data up to a 

%Let us focus on the region constrained by the data, where the surface brightness of each galaxy is higher than the background noise level (shown as a dotted horizontal line in the top panel). 
%The residual between the observed data and the flux predicted by the best-fit S\'{e}rsic model is shown at the bottom of each subpanel. In the region of 
%

\subsection{Stellar mass profile measurements}\label{ssec:mstar}

The stellar mass profiles of the galaxies in the sample were obtained by fitting stellar population synthesis models to the observed multi-band surface brightness profiles, following the same procedure adopted by \citet{SWB19}.
In summary, I assumed that the stellar component of each galaxy can be described by a composite stellar population (CSP) model with an exponentially declining star formation history, a single value for the metallicity $Z$ and a single value for the dust attenuation $\tau_V$, constant throughout the galaxy.
I then used the software {\sc Galaxev} \citep{B+C03} to predict, for each galaxy, broad-band magnitudes for each set of values of the stellar population parameters, including the stellar mass $M_*$, using semi-empirical stellar spectra from the BaSeL 3.1 library \citep{Wes++02}, Padova 1994 evolutionary tracks \citep{Fag++94a,Fag++94b,Fag++94c} and a Chabrier stellar initial mass function \citep[IMF][]{Cha03}.
These synthetic magnitudes were then fitted to the observed magnitudes in $g, r, i, z, y$ bands with a Markov Chain Monte Carlo (MCMC), using the python implementation {\sc emcee} \citep{For++13} of the affine-invariant sampling method introduced by \citet{G+W10}.
I assumed flat priors on $\log{M_*}$, on the stellar age (i.e. the time since the initial burst) and the star formation rate decay time, on $\log{\tau_V}$, and a Gaussian prior on $\log{Z}$ with a mean that scales with $\log{M_*}$, following the mass-metallicity relation of \citet{Gal++05}.
Following \citet{SWB19}, I added in quadrature a $0.05$ uncertainty to the observed magnitudes, to mimic the effect of systematic errors in the synthetic stellar population models: the latter are typically unable to fit observed magnitudes down to the (very small) photometric noise level. If not corrected for, this would result in the uncertainties on the stellar mass being underestimated \citep[see also Figure 3 of][and related discussion]{Son++19}.

The nominal output of the fitting procedure described above is the observed stellar mass $\mobs$ and related uncertainty, $\epsilon_*$, obtained by approximating the marginal posterior probability distribution function in $\log{M_*}$ as a Gaussian.
Then, self-consistently with the assumption of spatially constant colours and stellar population parameters, I assumed that the stellar mass density profile of a galaxy is described by a S\'{e}rsic profile with the same structural parameters inferred in the photometry fitting procedure and constant stellar mass-to-light ratio.

\section{The distribution in $\mten$-$\slope$ space}\label{sect:m10slope}

\begin{figure}
\includegraphics[width=\columnwidth]{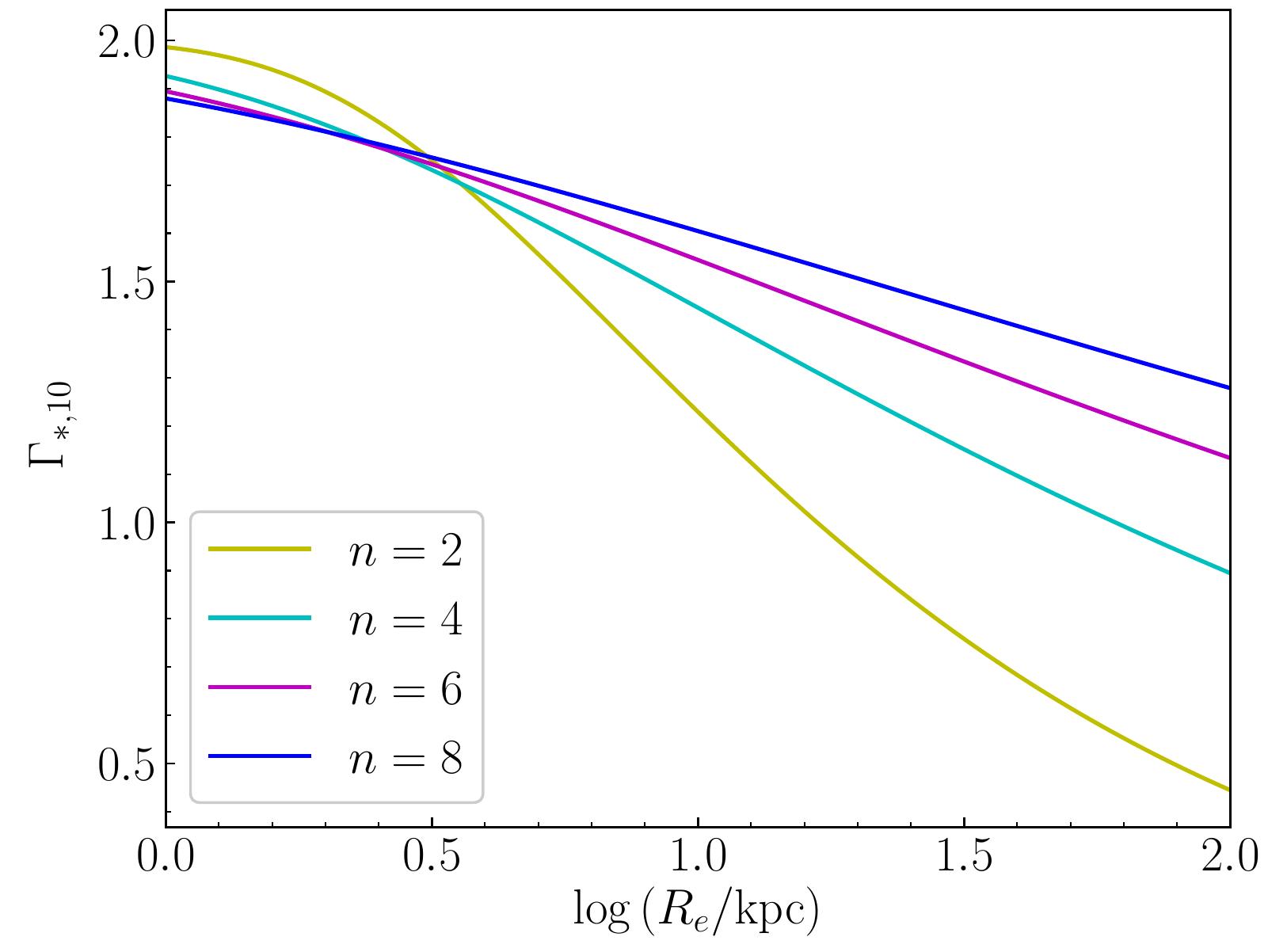}
\caption{
Mass-weighted slope of the stellar surface density profile within 10~kpc, defined in \Eref{eq:slope}, as a function of half-light radius, for a S\'{e}rsic profile, for four different values of the S\'{e}rsic index $n$.
\label{fig:slope_vs_reff}
}
\end{figure}

Before calculating the values of $\mten$ and $\slope$ for the galaxies in the sample, I first computed them for some example cases, in order to develop intuition on how this new parameterisation relates to quantities we are more used to, such as $\mstar$ and $\reff$.
I first point out that, under the assumption of a spatially constant mass-to-light ratio $M_*/L$, the mass-weighted slope is purely a function of the surface brightness profile $I(R)$: this can be seen by writing $\Sigma_*(R)=M_*/LI(R)$ in \Eref{eq:slope}, where the mass-to-light ratio cancels out.
Secondly, the maximum allowed value of the mass-weighted slope is $\slope=2$, which is obtained when the projected stellar mass density at $10$~kpc is zero.
Independently of the stellar mass of a galaxy, we can then ask how $\slope$ varies as a function of $\reff$ and $n$, for a S\'{e}rsic surface brightness profile. This is shown in \Fref{fig:slope_vs_reff} for a few values of $n$. The general trend is that of a decreasing slope for increasing $\reff$: if the half-light radius is large, the average surface brightness profile within 10~kpc is shallower.

\begin{figure}
\includegraphics[width=\columnwidth]{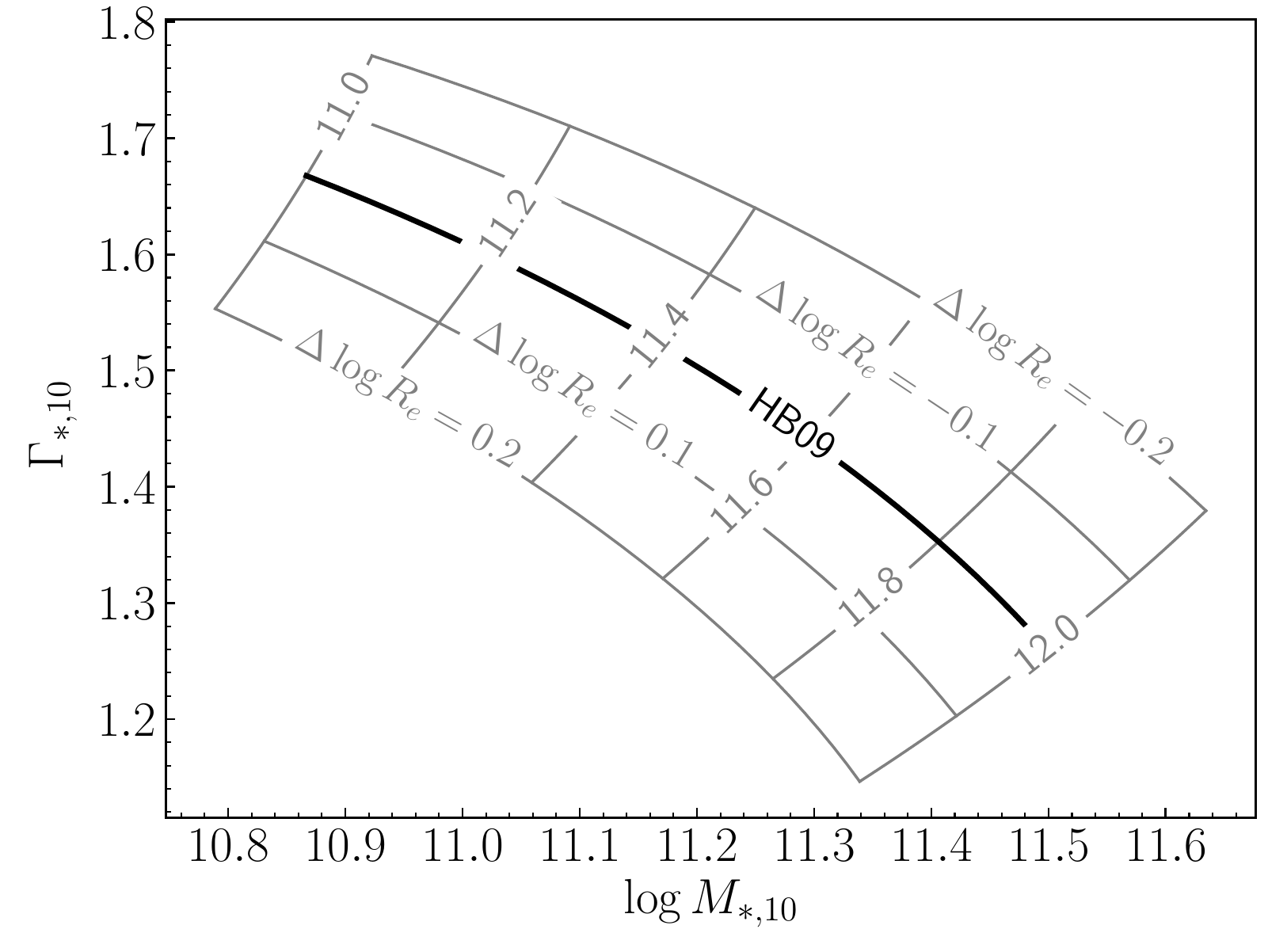}
\caption{
Distribution in $\mten-\slope$ space of galaxies following the average mass-size relation from HB09 (thick line).
Lines parallel to it refer to galaxies offset from the average by a fixed amount in $\log{\reff}$, as described on the labels.
Lines perpendicular to the HB09 relation are obtained by varying the half-light radius at fixed total stellar mass (the numbers shown on each line are the corresponding values of $\log{\mstar}$).
\label{fig:HB09}
}
\end{figure}

I then calculated the distribution in $(\mten,\slope)$ space of galaxies following a reference mass-size relation from the literature. For this purpose, I used the measurement of \citet[][BH09 from now on]{H+B09}, who fitted a quadratic relation to values of $\reff$ and $\mstar$ of SDSS early-type galaxies, obtained by modeling the stellar profile with a de Vaucouleurs model (a S\'{e}rsic profile with $n=4$).
The assumption of a de Vaucouleurs profile means that the stellar profile of galaxies in the HB09 sample constitute a 2-parameter family of objects, which can be easily mapped to the $(\mten,\slope)$ space.
In \Fref{fig:HB09}, I show the $\mten-\slope$ relation of galaxies lying on the average mass-size relation from HB09 (thick line). I also show lines of constant stellar mass and varying radius, as well as lines offset from the average mass-size relation by a fixed amount in $\log{\reff}$.
The mass enclosed within $10$~kpc increases with total stellar mass, while the projected slope decreases. This matches the expectation suggested by the plot in \Fref{fig:slope_vs_reff}: since more massive galaxies have a larger value of $\reff$, their stellar profile within $10$~kpc is shallower.
Lines of varying $\reff$ at fixed $\mstar$ are also shown in \Fref{fig:HB09}: galaxies that appear to be more compact (i.e. that lie below the mass-size relation), have higher values of $\mten$ and a steeper stellar slope. I point out, however, that this is a consequence of the assumption of a de Vaucouleurs profile and is not true in general: for instance, if one adopts the more general S\'{e}rsic profile and if S\'{e}rsic index correlates with stellar mass, a qualitatively different behaviour could emerge.

\begin{figure}
\includegraphics[width=\columnwidth]{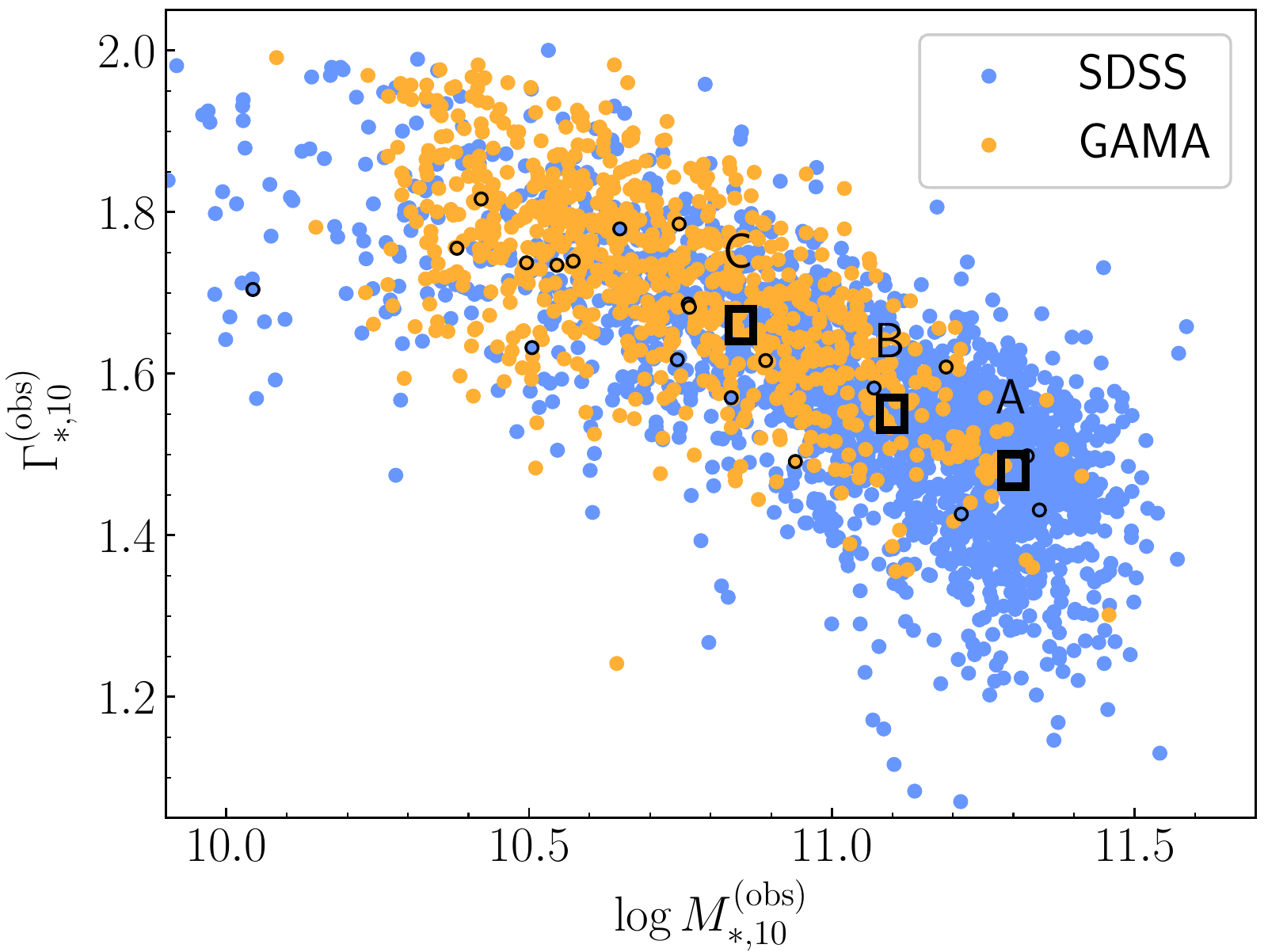}
\caption{Distribution in observed values of the stellar mass enclosed within $10$~kpc, $\mten$, and the mass-weighted stellar density slope within the same aperture, $\slope$, for the galaxies in the SDSS and GAMA samples. %Empty circles are data points flagged as outliers (typically associated with unrealistically large half-light radii) and removed from the sample.
%The color of each point indicates the galaxy redshift.
The three boxes are bins in $(\mten,\slope)$ space: the stellar density profile of the galaxies in each bin are shown in \Fref{fig:spread}.
\label{fig:m10_slope}
}
\end{figure}

In \Fref{fig:m10_slope}, I plot the distribution in $\mten$ and $\slope$ of the galaxies in the two samples. %, color-coded by redshift.
Since these are observed values, subject to errors, I indicate them as $\mtenobs$ and $\slopeobs$ respectively, in order to keep a clear distinction between true and observed quantities.
The data points cover a similar region as that spanned by the HB09 mass-size relation. This is an important consistency check, as there is substantial overlap between the HB09 sample and the SDSS galaxies considered in this work.
%A few galaxies appear to be outliers with respect to the main cloud of points: when inspected, these have typically very large measured half-light radii and I suspect the S\'{e}rsic fit to be unreliable for them. For this reason, I removed them from the sample with a sigma-clipping technique: I iteratively fitted a linear relation between $\log{\mtenobs}$ and $\slopeobs$, then removed from the sample any objects that are more than $5\sigma$ away from the relation, with $\sigma$ being the scatter in $\slopeobs$ at fixed $\mtenobs$.
%Outliers removed in this way are shown as empty circles in \Fref{fig:m10_slope} (a dozen more outliers are outside of the boundaries of the plot).
%

\begin{figure*}
\includegraphics[width=\textwidth]{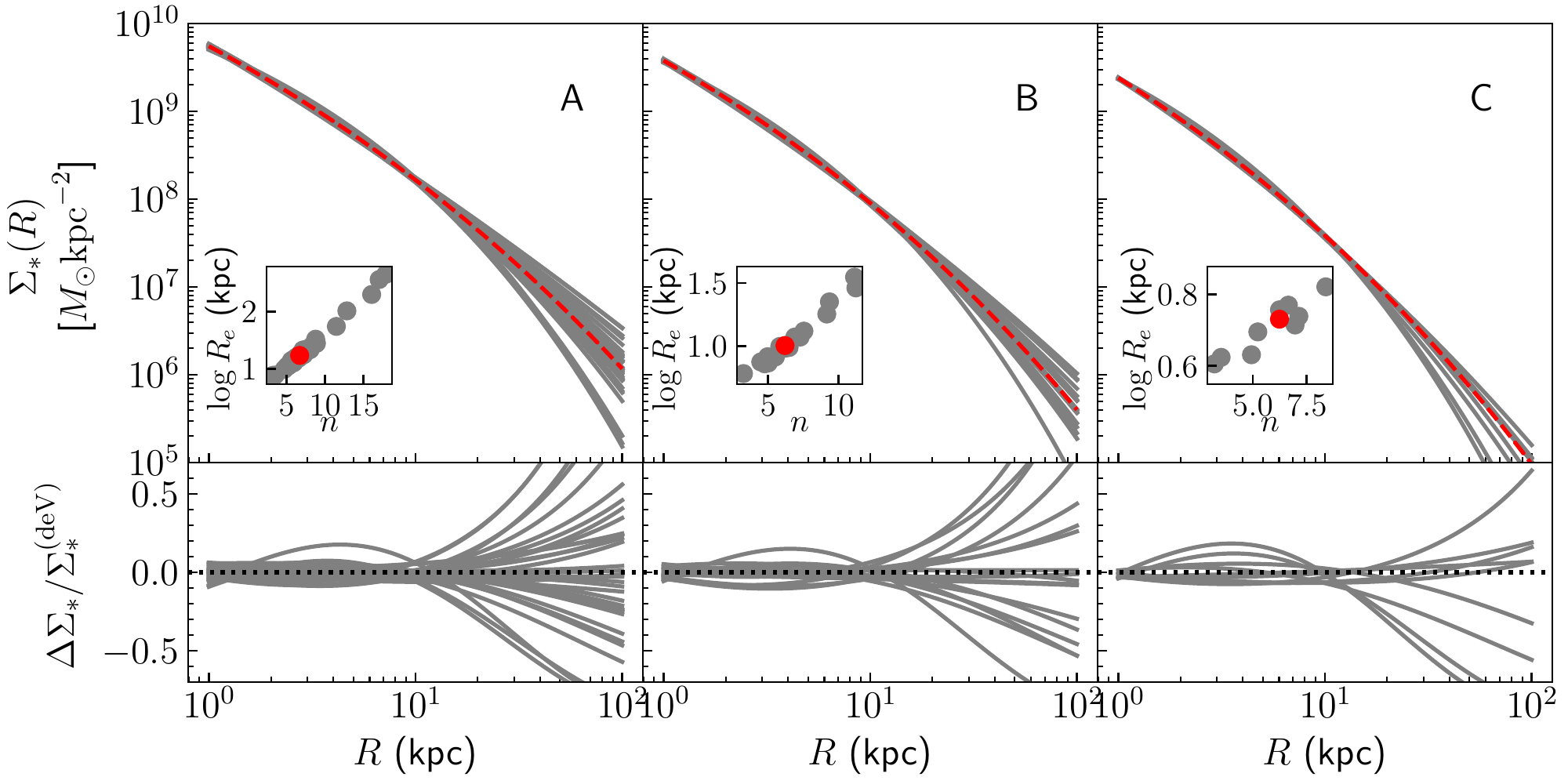}
\caption{
{\em Top:} stellar density profile, as inferred in subsection \ref{ssec:mstar}, of the galaxies in each of the three bins in $(\mten,\slope)$ shown in \Fref{fig:m10_slope}.
%The red dashed line is the stellar density profile of a de Vaucouleurs profile with the same value of $\mten$ and $\slope$ as the center of the bin.
The red dashed line is the stellar density profile of a S\'{e}rsic profile with the same index $n$ as the median among galaxies in the bin, and the same value of $\mten$ and $\slope$ as the center of the bin.
The inset shows the distribution in S\'{e}rsic index and half-light radius of each galaxy.
{\em Bottom:} residual between the stellar mass density of each galaxy in the bin and that of the reference profile with the same $\mten$ and $\slope$ as the center of the bin, divided by the latter.
\label{fig:spread}
}
\end{figure*}

The main motivation for introducing this new pair of variables $(\mten,\slope)$ is to use them to characterise the stellar density profile of elliptical galaxies in place of the more traditional $(\mstar,\reff)$ parameterisation.
This new parameterisation will be more useful the smaller the spread in stellar profiles of observed galaxies at fixed $(\mten,\slope)$.
In order to verify to what extent $(\mten,\slope)$ is a good predictor of the stellar density profile $\Sigma_*(R)$, I made three narrow bins in $(\mten,\slope)$ space, shown in \Fref{fig:m10_slope}, then plotted $\Sigma_*(R)$ of the galaxies in each bin in \Fref{fig:spread} (grey curves).
%Additionally, I considered as a reference the stellar density profile of a de Vaucouleurs model corresponding to the values of $\mten$ and $\slope$ of the bin center (red dashed line).
Additionally, I considered as a reference the stellar density profile of a S\'{e}rsic model with index $n$ equal to the median value among the galaxies in the bin, and same values of $\mten$ and $\slope$ of the bin center (red dashed line).
The galaxies in each bin span broad ranges of half-light radius and S\'{e}rsic index, as shown in the insets of \Fref{fig:spread}, yet they appear to share similar stellar profiles, particularly in the region $R<10\,{\rm kpc}$.
The stellar profiles start diverging at large radii, but those are also the regions poorly constrained by the data and where the S\'{e}rsic profile approximation is less accurate (see \Fref{fig:gof}).
The bottom panels of \Fref{fig:spread}, which show the relative differences in $\Sigma_*$ with respect to the reference profile, offer a more quantitative view: in the inner $10$~kpc the spread in stellar density is smaller than 20\%, with part of this spread being due to the finite width of each bin.
This means that, given the value of $\mten$ and $\slope$ of an elliptical galaxy, we can guess its inner stellar density profile with an accuracy of 20\% or better, by simply considering a S\'{e}rsic profile with the same values of $\mten$ and $\slope$.

\section{Velocity dispersion as a function of $\mten$ and $\slope$}\label{sect:veldisp}

%Another way of assessing how well the $\mten-\slope$ parameterization can summarise the properties of an elliptical galaxy is by 
The stellar profile of an elliptical galaxy is known to correlate with its dynamical state: elliptical galaxies are distributed around a thin surface in the three-dimensional space defined by central surface brightness (or central stellar mass density), half-light radius and central velocity dispersion, known as the fundamental plane \citep{Dre++87,D+D87,H+B09}.
Another way of assessing how well the $\mten-\slope$ parameterisation can summarise the properties of an elliptical galaxy is then by measuring how well it can predict its stellar velocity dispersion.
In this section I investigate the fundamental plane in the context of my new parameterisation of the stellar density profile: I measure how the central velocity dispersion is distributed as a function of $\mten$ and $\slope$, with a particular focus on the amplitude of the intrinsic scatter in velocity dispersion at fixed $(\mten,\slope)$ (i.e. the tightness of the relation).
%One should then expect the stellar velocity dispersion to also correlate with $\mten$ and $\slope$. In this section, I will investigate to what extent is $(\mten,\slope)$ a good predictor of the central velocity dispersion.

\begin{figure}
\includegraphics[width=\columnwidth]{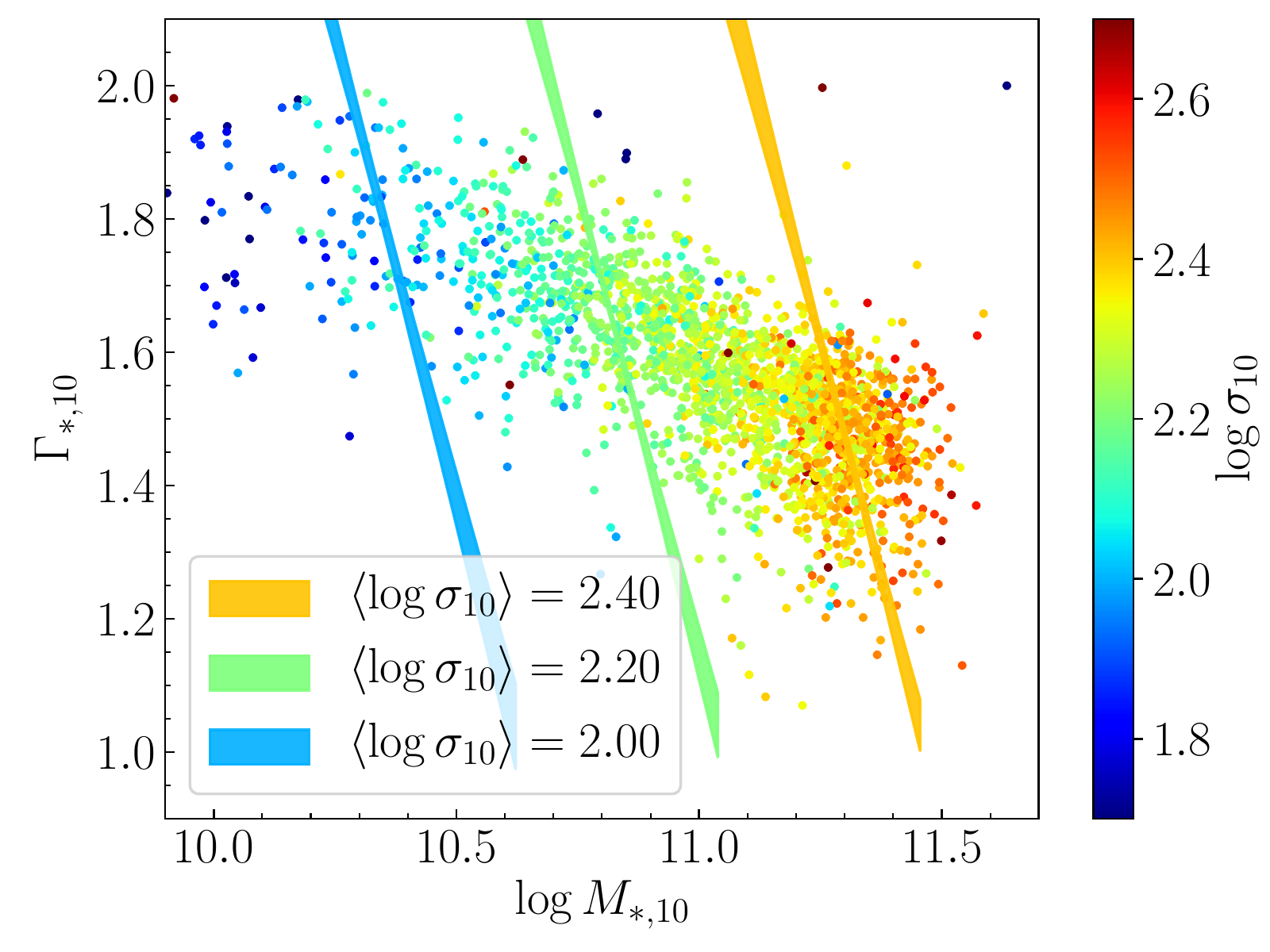}
\caption{
Distribution in $\mten-\slope$ space of the galaxies in the SDSS sample, colour-coded by stellar velocity dispersion within an aperture of $10$~kpc.
The three bands mark 68\% credible regions of values of the average stellar velocity dispersion, as inferred from the fit to the $\mten-\slope$ fundamental plane (\Eref{eq:modelsigmaten}). 
}
\label{fig:sigmasamp}
\end{figure}

For each galaxy in the SDSS sample, its line-of-sight stellar velocity dispersion measured within a circular fiber of the SDSS spectrograph, $\sigmaap$, is available.
In fundamental plane studies, this is usually converted into a central velocity dispersion, measured within a circular aperture with size equal to a fixed fraction of the half-light radius.
Since $\mten$ and $\sigma$ are calculated within a fixed physical aperture of radius $10$~kpc, it is more appropriate to consider the velocity dispersion within this aperture instead, labeled as $\sigmaten$. 
I obtained $\sigmaten$ by assuming the following radial scaling of the stellar velocity dispersion with aperture size, from \citet{Cap++06}:
\begin{equation}\label{eq:apcorr}
\sigmaten = \sigmaap \times \left(\frac{10\rm{kpc}}{\rap}\right)^{-0.066},
\end{equation}
where $\rap$ is the radius of the spectroscopic aperture, $1.5''$, in physical units at the redshift of the galaxy.
The distribution of SDSS galaxies in $\mten-\slope$ space is shown again in \Fref{fig:sigmasamp}, now colour-coded by the values of $\sigmaten$.

I fitted for the distribution of the galaxies in the SDSS sample in $(\mten,\slope,\sigmaten)$ space with a Bayesian hierarchical approach \citep[see e.g.][]{S+L18}. Each galaxy is described by its true values of $(\mten,\slope,\sigma)$: these are quantities at the bottom level of a hierarchy of parameters, drawn from a distribution describing the sample population, in turn described by parameters (usually referred to as hyper-parameters) that we wish to infer.
The main ingredient of the model in this case is the distribution of $\sigmaten$ as a function of $\mten$ and $\slope$. I considered the following Gaussian distribution in $\log{\sigmaten}$:
\begin{equation}\label{eq:modelsigmaten}
\log{\sigmaten}\sim\mathcal{N}(\mu_\sigma(\mten,\slope), s_\sigma^2),
\end{equation}
%\begin{equation}
%\pr(\sigmaten|\mten,\slope) = \frac{1}{\sqrt{2\pi}s_\sigma}\exp{\left\{-\frac{(\log{\sigmaten} - \mu_{\sigma})^2}{2s_\sigma^2}\right\}},
%\end{equation}
with a mean that allows for a linear scaling with $\log{\mten}$ and $\slope$,
\begin{equation}
\mu_{\sigma}(\mten,\slope) = \mu_{\sigma,0} + \beta (\log{\mten} - 11) + \xi (\slope - 1.6),
\end{equation}
and dispersion $s_\sigma$.
The hyper-parameters of this model are
\begin{equation}
\left\{\mu_{\sigma,0},\beta_\sigma,\xi_\sigma,s_\sigma\right\}.
\end{equation}
I sampled the posterior probability distribution of these hyper-parameters given the data, while marginalising over the parameters describing the individual (true) values of $\mten,\slope,\sigmaten$ of each galaxy, using an MCMC.
\Tref{tab:sigma} lists the inferred median and 68\% credible intervals of the marginal posterior distribution of each hyper-parameter.
The velocity dispersion correlates with the stellar mass enclosed within 10~kpc, $\beta_\sigma = 0.478\pm0.010$, as well as on the stellar density slope, $\xi_\sigma=0.17\pm0.02$.
Bands of constant average velocity dispersion in $\mten-\slope$ space are shown in \Fref{fig:sigmasamp}.
\begin{table*}
\caption{Inference on the hyper-parameters describing the distribution of the stellar velocity dispersion $\sigmaten$ in $(\mten,\slope)$ space. For each parameter, the median and 68\% credible interval of its marginal posterior PDF is reported.}
\label{tab:sigma}
\begin{tabular}{lcl}
\hline
\hline
Parameter & Med.$\pm 1\sigma$ & Description \\
\hline
$\mu_{\sigma,0}$ & $2.277 \pm 0.002$ & Mean $\log{\sigmaten}$ at $\mten=11.2$ and $\slope=1.5$. \\
$\beta_{\sigma}$ & $0.478 \pm 0.010$ & Dependence of $\log{\sigmaten}$ on $\mten$. \\
$\xi_{\sigma}$ & $0.17 \pm 0.02$ & Dependence of $\log{\sigmaten}$ on $\slope$. \\
$s_{\sigma}$ & $0.045 \pm 0.002$ & Scatter in $\log{\sigmaten}$ around the mean. \\

\end{tabular}
\end{table*}

\begin{figure}
\includegraphics[width=\columnwidth]{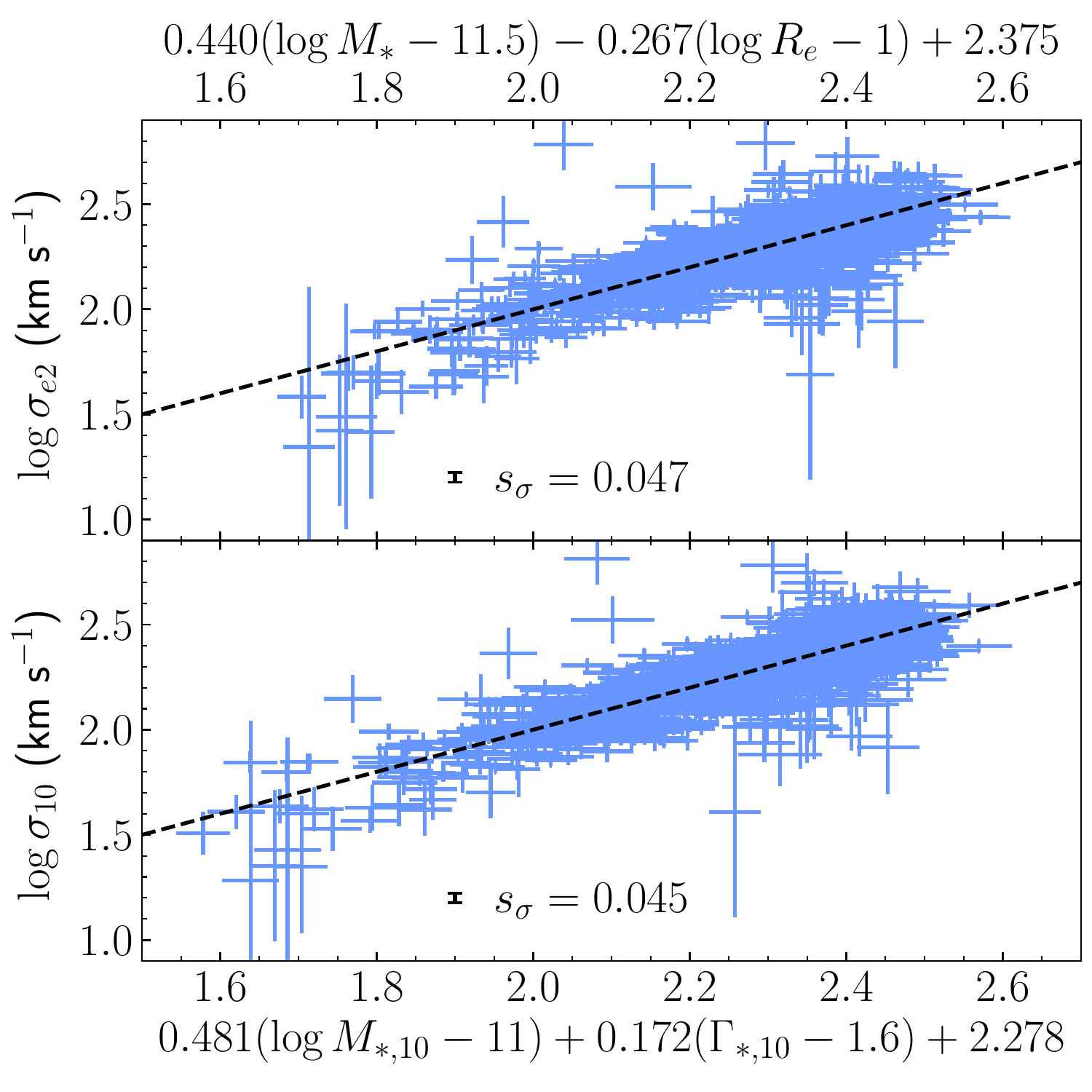}
\caption{
{\em Bottom}: stellar velocity dispersion within an aperture of $10$~kpc of the SDSS galaxies as a function of the inferred average $\log{\sigmaten}$ given the observed values of $(\mten,\slope)$, referred to in this paper as the $\mten-\slope$ fundamental plane.
{\em Top}: stellar velocity dispersion within an aperture of half the half-light radius as a function of the inferred average $\log{\sigmaetwo}$ given the observed values of $(\mstar,\reff)$, also known as the stellar mass fundamental plane.
The black error bar at the bottom of each panel shows the amplitude of the inferred intrinsic scatter around each relation.
\label{fig:fp}
}
\end{figure}

The $\mten-\slope$ fundamental plane is shown in the bottom panel of \Fref{fig:fp}.
The intrinsic scatter in $\log{\sigmaten}$ is $s_\sigma = 0.045\pm0.002$. This is $\sim30\%$ smaller than the scatter around the velocity dispersion-stellar mass relation, measured to be $0.066$ by \citet{CSN20} on a similarly selected sample of SDSS galaxies as those considered here.
It is interesting to compare the value of the intrinsic scatter to that measured around the stellar mass fundamental plane, which I constrained as follows.
I first obtained for each galaxy the stellar velocity dispersion within an aperture equal to half of its half-light radius, $\sigmaetwo$, using the same aperture correction expression of \Eref{eq:apcorr}, then fitted for the distribution of $\log{\sigmaetwo}$ as a function of $\mstar$ and $\reff$.
Similarly to the model of \Eref{eq:modelsigmaten}, I assumed a Gaussian distribution in $\log{\sigmaetwo}$. Using the same inference method described above, I found the following constraint for the mean of this Gaussian,
\begin{equation}
\mu_\sigma = 2.375 + 0.441(\log{\mstar}-11.5) - 0.268(\log{\reff} - 1),
\end{equation}
with errors on the coefficients similar in value to those of the previous model, and a dispersion of $s_\sigma = 0.047\pm0.002$.
This stellar mass fundamental plane is shown in the top panel of \Fref{fig:fp}.
The intrinsic scatter around the $\mten-\slope$ fundamental plane is the same in value as that around the stellar mass fundamental plane.
This means that $(\mten,\slope)$ is as good a predictor of the stellar velocity dispersion as the combination of stellar mass and half-light radius, with the advantage that $\mten$ and $\slope$ can be measured more robustly.

\section{Application: comparison with hydrodynamical simulations}\label{sect:eagle}

As an example of a possible use of this new description of the stellar surface mass density of elliptical galaxies, I show in this section a comparison between the $\mten-\slope$ relation of observed galaxies with that of simulated galaxies from the \eagle\ hydrodynamical simulations.
The stellar density profile cannot be considered a pure prediction of \eagle, because those simulations were partially tuned to avoid the creation of overly-compact galaxies. Nevertheless, checking the stellar distribution of simulated galaxies against observed ones can provide useful insight on the accuracy of some of the recipes adopted by \eagle, for instance the subgrid modeling of feedback from star formation and from active galactic nuclei.
\citet{Sch++15} compared the stellar mass-half-mass radius of \eagle\ galaxies with S\'{e}rsic index $\nser < 2.5$ against the stellar mass-half-light radius of observed galaxies from different studies (see their Figure 9).
They found that the simulations and the data agree to within $0.1$~dex, but were unable to draw stronger conclusions, partly because of inconsistencies in the definition of size between the different samples.
The adoption of a more robust definition of the stellar profile, based on $(\mten,\slope)$, should alleviate this issue.

\begin{figure}
\includegraphics[width=\columnwidth]{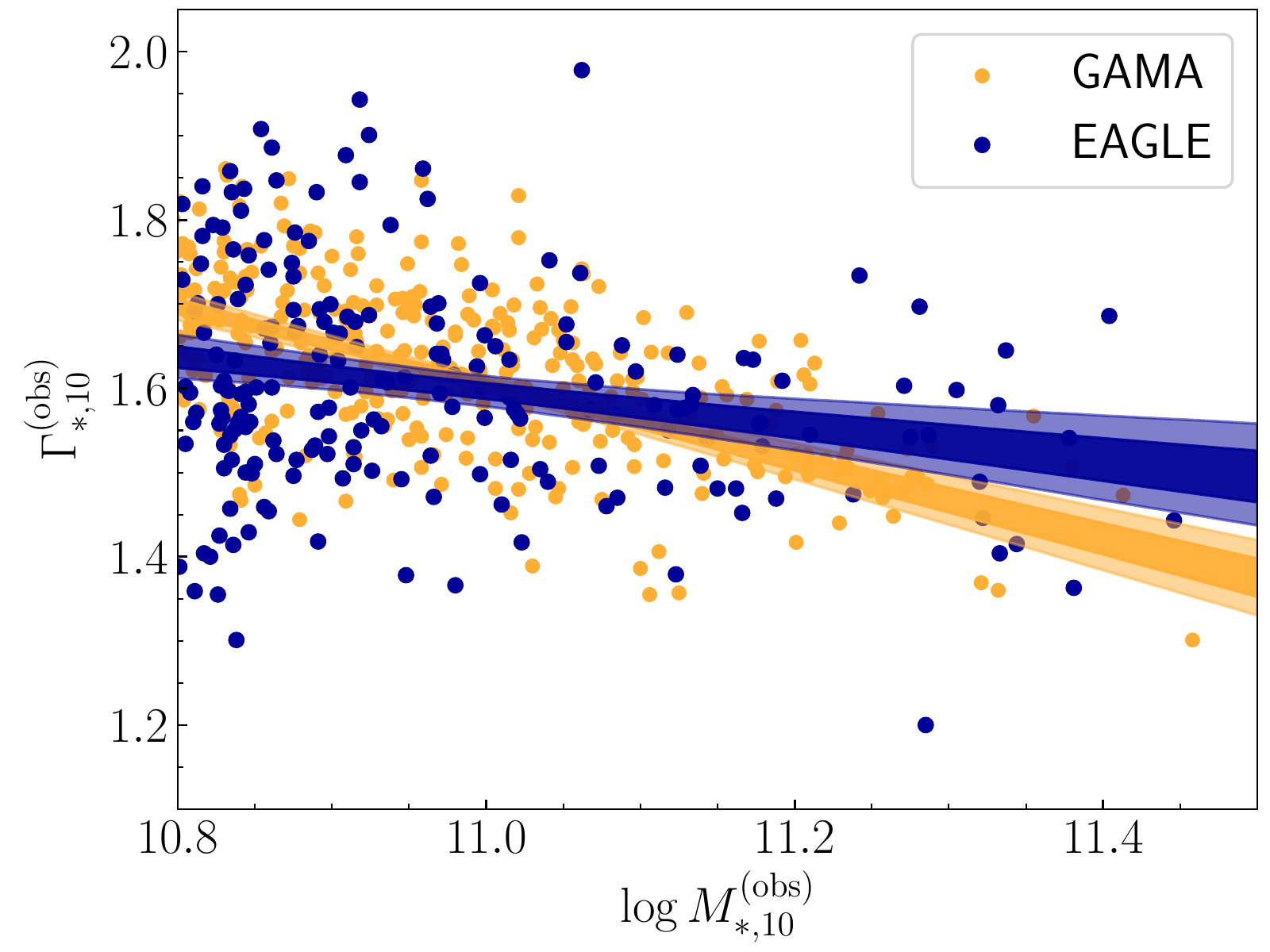}
\caption{Relation between $\mten$ and $\slope$ of a volume limited subsample of GAMA galaxies, and of similarly selected galaxies from the \eagle\ simulation.
The bands show the 68\% and 95\% credible regions of the $\mten-\slope$ relation obtained by fitting the model of \Eref{eq:slopedist} to each sample.
\label{fig:eagle}
}
\end{figure}

%I wish to compare the distribution in the mass-weighted projected slope $\slope$ at fixed $\mten$ of observed elliptical galaxies with that of similarly selected galaxies from EAGLE.
In order to measure the $\mten-\slope$ relation, it is necessary to work with a sample of galaxies for which the completeness in $\mten$ is independent of the slope $\slope$ (or to apply a $\slope$-dependent incompleteness correction). A volume limited sample, complete down to a minimum stellar mass, satisfies this condition (though is not strictly necessary).
I obtained such a sample by considering only galaxies from the GAMA sample, and by applying the following cut to the observed stellar masses within $10$~kpc: $\log{\mtenobs} > 10.80$. As shown in Appendix~\ref{sect:appendixa}, the resulting sample is more than 95\% complete in $\mtenobs$.
The distribution in $\mten-\slope$ space of this sample is shown in \Fref{fig:eagle}.

I then selected elliptical galaxies from the $z=0.1$ snapshot of the \eagle\ simulation {\sc RefL0100N1504}, which corresponds to the {\sc EAGLE Reference} model and the largest box size (100~cMpc on the side).
I first considered all subhalos with a total stellar mass larger than $10^{10.8}M_\odot$, then measured $\mten$ and $\slope$ from the particle data. This was done by projecting particle positions along the third axis, estimating the axis ratio and position angle of the projected stellar mass distribution from its second moments, defining circularised coordinates using \Eref{eq:ellcoord}, calculating $\Sigma_*(10)$ on a thin elliptical annulus around a circularised radius of 10~kpc, obtaining $\mten$ by summing the stellar mass within the same radius and calculating $\slope$ using \Eref{eq:slope}.
Another advantage of using $(\mten,\slope)$, is the fact that both quantities are defined at a sufficiently large radius to be immune from finite resolution effects in the simulation (the gravitational softening length in the large \eagle\ box is $2.66$ comoving kpc). This is not the case when trying to estimate the half light radius of a simulated galaxy by fitting a S\'{e}rsic surface brightness profile to it: since the inner regions are artificially smoothed out, the inferred stellar density profile will be biased.

I applied a cut to select only galaxies with $\log{\mten} > 10.80$, as done with the observed GAMA sample, which reduced the sample size to 316, and then visually inspected SDSS-like RGB images, produced by \citet{Tra++17}, of the remaining objects, to select only galaxies with elliptical morphology.
With this visual selection step, 111 spiral or irregular galaxies were removed, bringing the final size of the \eagle\ sample to 205 objects.
I also explored an alternative selection, based on a cut in the fraction of the kinetic energy in ordered corotation, $\kappa_{co} < 0.4$, as suggested by \citet{CST19}, with negligible changes in the outcome of the analysis presented in the rest of this section.

I fitted for the distribution in $\slope$ as a function of $\mten$ of both the GAMA and the \eagle\ samples. I modeled this as a Gaussian in $\slope$
\begin{equation}\label{eq:slopedist}
\slope \sim \mathcal{N}(\mu_{\Gamma}(\mten), s_\Gamma^2),
\end{equation}
with a mean that scales linearly with $\log{\mten}$,
\begin{equation}\label{eq:mtensloperel}
\mu_\Gamma(\mten) = \mu_{\Gamma,0} + \beta_\Gamma (\log{\mten} - 11),
\end{equation}
and dispersion $s_\Gamma$.
I fitted for the model parameters $\mu_{\Gamma,0},\,\beta_\Gamma,\,s_\Gamma$ with a Bayesian hierarchical inference method. In the case of the fit to the observed data, this was done by marginalising over the parameters describing the individual values of the stellar mass of each galaxy, $\mten$, given its observed value, $\mtenobs$, and correcting for the Eddington bias introduced by the cut in $\mtenobs$, following \citet{S+L18}.
The median and 68\% credible interval of the marginal posterior probability distribution of each parameter is given in \Tref{tab:inference}, and the 68\% and 95\% credible regions for the average value of the stellar slope, $\mu_\Gamma$, is plotted in \Fref{fig:eagle} as a function of $\mten$.
\begin{table*}
\caption{Inference on the hyper-parameters describing the distribution of the stellar density slope $\slope$ as a function of the stellar mass enclosed within 10~kpc, $\mten$, for the GAMA and \eagle\ samples. For each parameter, the median and 68\% credible interval of its marginal posterior PDF is reported.}
\label{tab:inference}
\begin{tabular}{lccl}
\hline
\hline
Parameter & GAMA & \eagle\ & Description \\
\hline
$\mu_{\Gamma,0}$ & $1.606 \pm 0.005$ & $1.597 \pm 0.009$ & Mean $\slope$ at $\mten=11.0$. \\
$\beta_{\Gamma}$ & $-0.46 \pm 0.04$ & $-0.20 \pm 0.06$ & Dependence of $\slope$ on $\mten$. \\
$s_{\Gamma}$ & $0.075 \pm 0.004$ & $0.127 \pm 0.006$ & Scatter in $\slope$ around the mean. \\

\end{tabular}
\end{table*}
%

%The mean of the $\slope$ distribution at the pivot point of \Eref{eq:mtensloperel}, $\log{\mten}=11.0$
The $\mten-\slope$ relation of the \eagle\ sample agrees very well with the observed one at the pivot point of \Eref{eq:mtensloperel}, $\log{\mten}=11.0$. However, the distribution of \eagle\ galaxies has a shallower slope and a larger intrinsic scatter compared to that of the GAMA sample.
At the low mass end, $\log{\mten}=10.80$, the maximum a posteriori value of the mean of the $\slope$ distribution of \eagle\ is $\sim0.06$ lower than the corresponding value of the observed sample. Since the observed intrinsic scatter in $\slope$ is $s_\Gamma=0.075\pm0.004$, the mean $\slope$ of \eagle\ galaxies, $\mu_\Gamma^{\mathrm{(EAGLE)}}$, is then roughly as shallow as that of the 20\%-ile of the observed distribution of GAMA galaxies.
Vice versa, towards the high mass end, $\log{\mten}=11.3$, the mean $\slope$ of \eagle\ galaxies is as steep as the 80\%-ile of the observed distribution.

\begin{figure}
\includegraphics[width=\columnwidth]{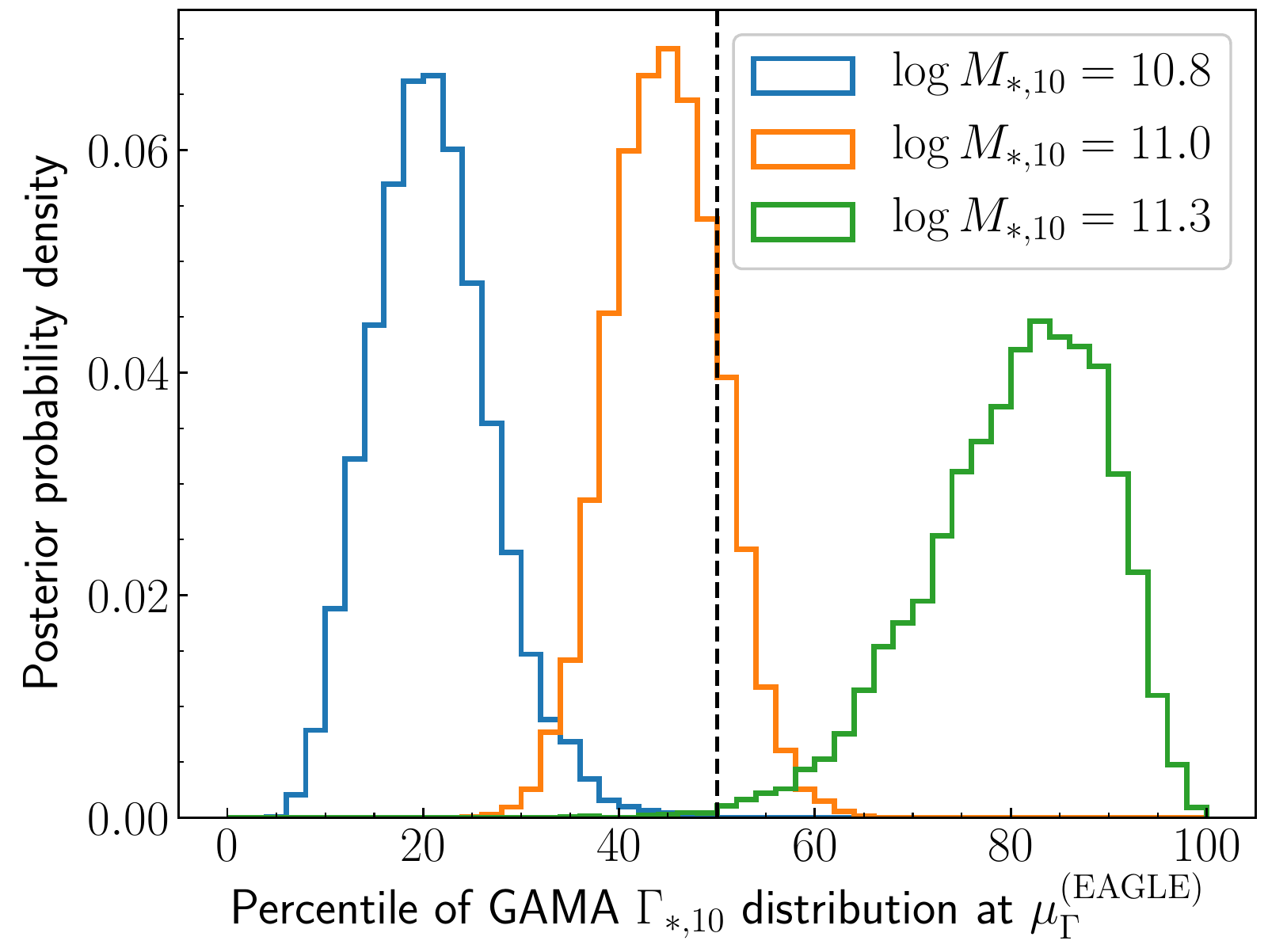}
\caption{Posterior predictive distributions of the percentile of the distribution in the stellar density slope of GAMA galaxies at $\slope=\mu_\Gamma^{\mathrm{(EAGLE)}}$, i.e. at the value of $\slope$ equal to the average of the \eagle\ sample, at three values of $\mten$.
The vertical dashed line marks the 50\%-ile, which is the expectation value in case of perfect agreement between the distribution of GAMA and \eagle\ galaxies.
\label{fig:pp}
}
\end{figure}

\Fref{fig:pp} shows a statistically more robust assessment of the difference between the two distributions, based on posterior prediction. For each set of values of the posterior probability distributions from the two inferences, and at three reference values of $\mten$, I first calculated the mean slope of the \eagle\ sample, $\mu_\Gamma^{\mathrm{(EAGLE)}}$, and then found the percentile of the distribution of the observed sample at $\slope=\mu_\Gamma^{\mathrm{(EAGLE)}}$. The three histograms in \Fref{fig:pp} show the distributions in percentiles obtained by marginalising over the posterior probability distributions, i.e. propagating the uncertainties.

The fact that the histogram relative to the $\log{\mten}=10.80$ point lies completely to the left of the 50\%-ile line means that, even when accounting for uncertainties, the stellar slope of the average \eagle\ galaxy is still shallower compared to observations at the same value of $\mten$. Conversely, the $\log{\mten}=11.30$ histogram lies to the right of the 50\%-ile line, indicating that at the high mass end \eagle\ galaxies have too steep of a stellar density slope.
This finding could be used to improve the physical prescriptions of the simulation, though that is beyond the goals of this work.

\section{Discussion}\label{sect:discuss}

\subsection{What is $(\mten,\slope)$ useful for?}

There are many definitions of the size of a galaxy in the scientific literature. In addition to the already discussed half-light radius, there are the Petrosian radius \citep{Pet76} and the Kron radius \citep{Kro80} \citep[see][for a discussion on how these relate to a S\'{e}rsic surface brightness profile]{G+D05}.
Then there are definitions based on the radius at which the surface brightness of a galaxy reaches a threshold value, such as the Holmberg radius \citep{Hol58}, or the more recent approach of \citet{Tru++20} based on a limiting stellar surface mass density.
All of these definitions are designed to capture most of the light of a galaxy, or a fixed fraction of it.
The approach of this work is radically different: by using $(\mten,\slope)$ to characterise the stellar density profile of a galaxy, one gives up quantifying the total stellar mass of a galaxy and focuses instead on a region well constrained by the data.

The rationale for the introduction of $(\mten,\slope)$ is to provide as observationally robust a way as possible to characterise the stellar profile of a galaxy with only two parameters.
Robustness is important when comparing observations with simulations, but also when doing comparisons between different observational datasets.
With most other definitions, the parameters used to describe the stellar profile of a galaxy will in general depend on the wavelength covered by the data, the depth of the data and, if the measurement involves fitting a model surface brightness profile, the adopted model family.
Even when the same data for the same galaxy is fitted with the same family of surface brightness models, for example a S\'{e}rsic profile, results can still vary depending on the adopted priors on the model parameters: many authors choose to limit the S\'{e}rsic index $\nser$ below a maximum allowed value (typically between 6 and 8), leading to different measurements of the total flux and half-light radius on galaxies for which the data favors larger values.

One caveat of the $(\mten,\slope)$ parameterisation is that, given the tilt of the distribution of galaxies in $\mten-\slope$ space, neither $\mten$ nor $\slope$ can be used in isolation to provide an insightful description of the stellar profile of a galaxy: both parameters are needed.
Conversely, the more traditional $(\mstar,\reff)$ approach might seem more appealing, because the total stellar mass is, with good reasons, generally regarded as a fundamental property of a galaxy.
That allows one to rank galaxies according to $\mstar$ and gain intuition on how the more massive objects differ from the least massive ones in terms of their assembly history and related properties.
At the same time, however, it is highly unlikely that a description based on a single parameter, the stellar mass, can be sufficient to capture all the complexities of galaxy evolution.
Even objects as apparently simple as elliptical galaxies exhibit nontrivial variations with half-light radius at fixed stellar mass, for example in their stellar population properties \citep{Dia++19} or in the mass of their central supermassive black hole \citep{vdB16}.
%A parameterization based purely on the stellar mass has therefore limited 
A description based on at least two parameters should therefore be preferable: although increasing the dimensionality of a problem introduces challenges in terms of data visualization and analysis, modeling choices should ultimately be dictated by the complexity of the physical system one wishes to describe.

The pair of values $(\mten,\slope)$ can replace $\mstar$ in all applications for which the latter is used merely as a label. Galaxies can still be ranked by identifying curves in $\mstar-\slope$ space that correspond to constant values of a variable of interest. For example, by moving perpendicularly with respect to the bands on \Fref{fig:sigmasamp} one finds galaxies with progressively different values of the velocity dispersion.

The stellar mass of a galaxy is often used in cosmology as a noisy proxy for the mass of its dark matter halo: this is for example the basic assumption of abundance matching approaches.
%Abundance matching can be still be done using $(\mten,\slope)$, as long 
%By combining $\mten$ and $\slope$ in such a way that the scatter in halo mass at fixed $(\mten
%Nothing prevents one to do abundance matching with $(\mten,\slope)$:
Abundance matching can still be done with $(\mten,\slope)$: it is sufficient to combine the two parameters in a new scalar variable (for example, a line in $\log{\mten}-\slope$ space) and rank galaxies according to it.
In fact, just like a certain combination of $\mten$ and $\slope$ can predict velocity dispersion with a smaller intrinsic scatter compared to stellar mass alone, it is plausible that another combination of the two parameters can provide a better proxy for halo mass, compared to the traditional abundance matching ansatz. Such combination could be determined with direct probes of halo mass, for example by measuring the weak gravitational lensing signal of a sample of galaxies as a function of $\mten$ and $\slope$.

Cases in which a measurement of the stellar profile at large radii is critical, however, will not benefit much from switching to a $(\mten,\slope)$ parameterisation, given the substantial spread in stellar profiles at $R>30\,{\rm kpc}$ at fixed $\mten$ and $\slope$ observed in \Fref{fig:spread}.
One such example is the determination of the fraction of total baryonic matter locked in stars: in this case, it is preferable to explicitly model the distribution of stellar mass out to the largest possible radii.
%In cases in which a measurement of the stellar profile at large radii is critical, however, such as an investigation of the fraction of total baryonic matter locked in stars, will not benefit much from switching to a $(\mten,\slope)$ parameterization, given the substantial spread in stellar profiles at $R>30\,{\rm kpc}$ at fixed $\mten$ and $\slope$ observed in \Fref{fig:spread}.
 
%A parameterization in terms of $(\mten,\slope)$ might seem less appealing than the more traditional $(\mstar,\reff)$ one, because neither $\mten$ nor $\slope$ can be used in isolation to 

%Indeed, there are cases for which a quantification of the total stellar mass of a galaxy is preferable, 
%Indeed, studies for which a measurement of the stellar profile at large radii is critical, such as investigations of the fraction of total baryonic matter locked in stars, will not benefit much from switching to a $(\mten,\slope)$ parameterization, given the substantial spread in stellar profiles at $R>30\,{\rm kpc}$ at fixed $\mten$ and $\slope$ observed in \Fref{fig:spread}.

%At the same time, studies for which a measurement of the stellar profile at large radii is critical, such as investigations of the fraction of total baryonic matter locked in stars, will not benefit much from switching to a $(\mten,\slope)$ parameterization, given the substantial spread in stellar profiles at $R>30\,{\rm kpc}$ at fixed $\mten$ and $\slope$ observed in \Fref{fig:spread}.

\subsection{Limitations and possible extensions}

The choice of 10~kpc as the reference scale within which to measure the enclosed stellar mass and the stellar density slope is clearly arbitrary.
It is a sensible choice for the type of objects studied in this work, elliptical galaxies, because they are typically massive and have sizes (in the half-light radius sense) on the same order of magnitude: in other words, an aperture of 10~kpc encloses a fraction of their total stellar mass that is substantial, yet smaller than unity.
The $\mten,\slope$ definition, however, becomes increasingly less useful at the low mass end: galaxies with a very small size have a correspondingly small value of the stellar surface mass density at $10$~kpc and consequently a mass-weighted stellar slope $\slope\approx2$. Dwarf galaxies then occupy a narrow horizontal band in $\mten-\slope$ space close to the maximum allowed value of $\slope=2$, meaning that what was originally intended as a two-parameter description of galaxies reduces effectively to a single parameter one, at low masses.
This limitation could be overcome by measuring the enclosed stellar mass and the mass-weighted slope at more than one radius, but that would lead to a much more cumbersome description compared to the single aperture case advocated in this work.

Although this work was focused entirely on elliptical galaxies, it is possible to measure $\mten,\slope$ of a galaxy of any morphology, provided that accurate measurements of its stellar profile out to at least 10~kpc is available. 
This is not a trivial requirement, especially for objects with multiple component such as disks, rings, spiral arms or bars. 
Additionally, while the value of $(\mten,\slope)$ of an elliptical galaxy gives us a precise idea of its stellar profile, at least in the region $R<10\,{\rm kpc}$, the same is not true for galaxies with a more complex morphology.
As an extreme case, we can consider the example of a galaxy consisting of a bulge and a star-forming ring at exactly $R=10\,{\rm kpc}$. Such an object would have a shallow stellar slope (i.e. a small value of $\slope$) in virtue of the high stellar mass density at $10\,{\rm kpc}$, but if we were to consider a S\'{e}rsic profile with the same values of $\mten$ and $\slope$, as done in \Fref{fig:spread}, the resulting stellar profile would be radically different from the true one, as it would be extending out to much larger radii.
That, however, is a general problem that arises when models with a small number of parameters are used to describe galaxies that are too complex: fitting a ring galaxy with a S\'{e}rsic profile, or even with a model consisting of the sum of a de Vaucouleurs profile and an exponential one, typically used to separate bulge and disks components, results usually in a poor fit and an overestimate of its stellar mass.

%What is $(\mten,\slope)$ useful for? Clearly, if measuring the stellar profile at large radii is critical, such as in studies aimed at determining the fraction of baryonic matter bound in stars, 

\subsection{Residual systematic uncertainties}

The main source of systematic uncertainty affecting measurements of $(\mten,\slope)$ as carried out in this work, is in the accuracy of the stellar population synthesis models used to convert measurements of the surface brightness in stellar surface mass density.
The estimate of the stellar mass-to-light ratio of a galaxy from spectral energy distribution fitting is subject to uncertainties on the stellar evolution models, stellar templates, choice of the stellar IMF, dust attenuation model, and parameterisation of the star formation history, to name a few.
%State-of-the art stellar population synthesis models are subject to uncertainties on the stellar evolution models, s
Especially when broad band photometric observations are used to constrain the stellar population parameters of a galaxy, one is usually forced to make a series of simplifying assumptions, as the data typically only offers a limited number of constraints.

In this work, I assumed spatially constant stellar population parameters. However, galaxies are known to have gradients in age and metallicity, which correspond in general to gradients in the stellar mass-to-light ratio.
This can be a problem when comparing values of $(\mten,\slope)$ between observations and simulations, unless measurements are carried out self-consistently in both realms.
This problem is not an exclusive of the $(\mten,\slope)$ approach, though, but applies more generally to any attempt at describing the stellar profile of a galaxy with a model that is too simple.
%It could be solved by allowing for spatial variations in stellar population parameters when converting multi-band surface brightness profile measurements into a stellar mass density profile
It could be solved by allowing for spatial variations in stellar population parameters in the model, provided that the data is sufficiently constraining.
Alternatively, one could more conservatively settle for measuring enclosed light and surface brightness density slope in place of $(\mten,\slope)$, leaving to the simulation side the burden of predicting the observed light distribution.

\section{Summary}\label{sect:concl}

I introduced a new parameterisation for the stellar mass density profile of elliptical galaxies, based on the projected stellar mass enclosed within an aperture of $10$~kpc, $\mten$, and the mass-weighted stellar density slope within $10$~kpc, $\slope$.
These two parameters can be measured more robustly than the total stellar mass and half light radius of a galaxy, because they do not rely on the detection of light at large radii and do not suffer from ambiguities related to the definition of the boundary of a galaxy.

I measured the distribution of $\mten$ and $\slope$ of two samples of elliptical galaxies at $z<0.5$ using photometric data from the HSC survey.
I found that, by specifying the values of $\mten$ and $\slope$ of an elliptical galaxy, it is possible to guess its stellar density profile in the inner $10$~kpc with better than 20\% accuracy.
I then investigated the distribution in stellar velocity dispersion as a function of $(\mten,\slope)$, dubbed the $\mten-\slope$ fundamental plane. At fixed $(\mten,\slope)$, the intrinsic scatter in stellar velocity dispersion is equal to that around the stellar mass fundamental plane.
This implies that no information on the dynamical mass of a galaxy is lost when adopting a parameterisation based on measurements carried out at a fixed aperture in place of $\mstar$ and $\reff$.

I then compared the distribution of $\slope$ as a function of $\mten$ of elliptical galaxies from the \eagle\ {\sc Reference} simulation with that of observed galaxies. The two parameters can be measured easily in simulated galaxies and are more immune to finite resolution effects compared to measurements based on the fit of a stellar density profile.
At $\mten=10^{11}M_\odot$, \eagle\ galaxies have the same average $\slope$ as measured in the observations. However, the \eagle\ $\mten-\slope$ relation is shallower and has a larger intrinsic scatter compared to observations. This result can be used as a clue to improve the accuracy of some of the physical prescriptions used in the simulation.

%In conclusion, although neither $\mten$ nor $\slope$ is designed to represent any fundamental property of a galaxy, this new description provides a robust way of differentiating among elliptical galaxies with different stellar profiles, and can be used to carry out accurate comparisons between the stellar profiles of observed and simulated galaxies.
In conclusion, although neither $\mten$ nor $\slope$ is designed to represent any fundamental property of a galaxy, this new description provides a robust way of differentiating among elliptical galaxies with different stellar profiles, and can be used to make accurate comparisons between stellar profiles of observed and simulated galaxies.
Combinations of $\mten$ and $\slope$ can substitute $\mstar$ in all applications for which a census of the total stellar mass of a galaxy is not a crucial requirement, such as cases in which the stellar mass is used as a proxy for a secondary quantity.

\begin{acknowledgements}
I thank Peter Mitchell, Matthieu Schaller, James Trayford and Nastasha Wijers for their help with \eagle\ data products, and Hendrik Hoekstra for his useful comments.
I acknowledge funding from the European Union's Horizon 2020 research and innovation programme under grant agreement No 792916.
The Hyper Suprime-Cam (HSC) collaboration includes the astronomical communities of Japan and Taiwan, and Princeton University.  The HSC instrumentation and software were developed by the National Astronomical Observatory of Japan (NAOJ), the Kavli Institute for the Physics and Mathematics of the Universe (Kavli IPMU), the University of Tokyo, the High Energy Accelerator Research Organization (KEK), the Academia Sinica Institute for Astronomy and Astrophysics in Taiwan (ASIAA), and Princeton University.  Funding was contributed by the FIRST programme from Japanese Cabinet Office, the Ministry of Education, Culture, Sports, Science, and Technology (MEXT), the Japan Society for the Promotion of Science (JSPS),  Japan Science and Technology Agency  (JST),  the Toray Science  Foundation, NAOJ, Kavli IPMU, KEK, ASIAA,  and Princeton University.
Funding for SDSS-III has been provided by the Alfred P. Sloan Foundation, the Participating Institutions, the National Science Foundation, and the U.S. Department of Energy Office of Science. The SDSS-III web site is http://www.sdss3.org/.
SDSS-III is managed by the Astrophysical Research Consortium for the Participating Institutions of the SDSS-III Collaboration including the University of Arizona, the Brazilian Participation Group, Brookhaven National Laboratory, Carnegie Mellon University, University of Florida, the French Participation Group, the German Participation Group, Harvard University, the Instituto de Astrofisica de Canarias, the Michigan State/Notre Dame/JINA Participation Group, Johns Hopkins University, Lawrence Berkeley National Laboratory, Max Planck Institute for Astrophysics, Max Planck Institute for Extraterrestrial Physics, New Mexico State University, New York University, Ohio State University, Pennsylvania State University, University of Portsmouth, Princeton University, the Spanish Participation Group, University of Tokyo, University of Utah, Vanderbilt University, University of Virginia, University of Washington, and Yale University.
GAMA is a joint European-Australasian project based around a spectroscopic campaign using the Anglo-Australian Telescope. The GAMA input catalogue is based on data taken from the Sloan Digital Sky Survey and the UKIRT Infrared Deep Sky Survey. Complementary imaging of the GAMA regions is being obtained by a number of independent survey programmes including GALEX MIS, VST KiDS, VISTA VIKING, WISE, Herschel-ATLAS, GMRT and ASKAP providing UV to radio coverage. GAMA is funded by the STFC (UK), the ARC (Australia), the AAO, and the participating institutions. The GAMA website is http://www.gama-survey.org/

\end{acknowledgements}

\bibliographystyle{aa}
\bibliography{references}

\appendix
%\onecolumn
\section{A nearly complete sample in $\mten$}\label{sect:appendixa}

To obtain a subset of galaxies approximately complete in $\mtenobs$, I first considered only objects from the GAMA sample. These were preselected to be in the redshift range $0.15 < z < 0.20$.
I then determined the minimum value of the observed stellar mass within $10$~kpc, $\mtenobsmin$, for which the GAMA sample is at least 95\% complete at $z=0.20$, and restricted the analysis to galaxies with $\mtenobs > \mtenobsmin$. 

\begin{figure}
\includegraphics[width=\columnwidth]{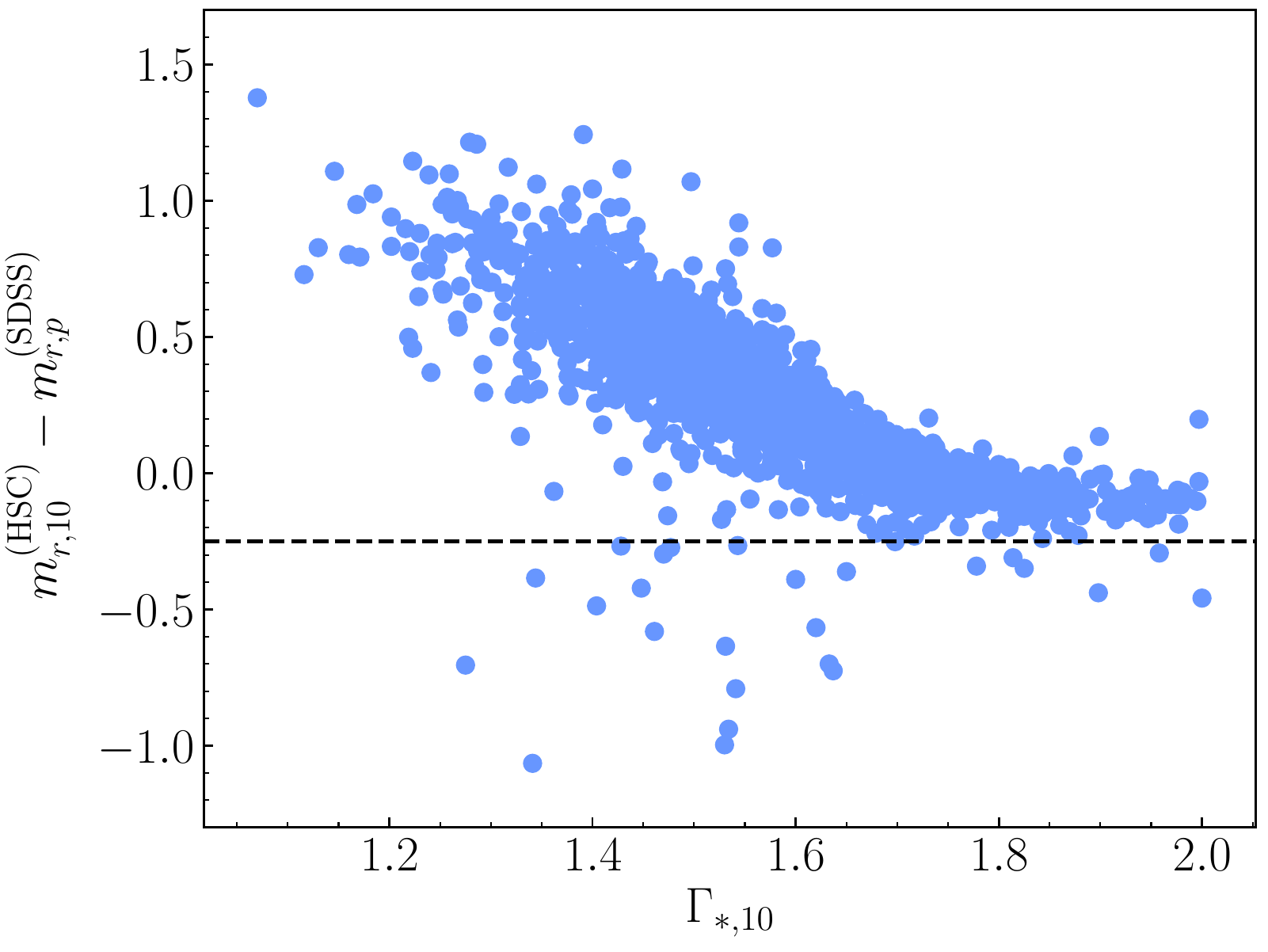}
\caption{
Difference between the HSC-based $r-$band magnitude within an aperture of $10$~kpc and the $r-$band Petrosian magnitude from SDSS DR12, as a function of the stellar density slope.
The horizontal line at $\mrten-\mrpetro=-0.25$ marks the assumed lower limit on the difference between the two magnitude measurements.
\label{fig:petro}
}
\end{figure}

The GAMA DR2 is complete down to an apparent $r-$band Petrosian magnitude $\mrpetro < 19.4$ in the field G15 (from which the galaxies used in this study are drawn), where $\mrpetro$ was obtained from the SDSS \citep[see][for the definition of Petrosian quantities used by the SDSS]{Bla++01,Yas++01}.
SDSS Petrosian magnitudes, however, map nontrivially to HSC magnitudes in the context of this work. \Fref{fig:petro} shows the difference between the HSC $r-$band magnitude calculated within an aperture of $10$~kpc, $\mrten$, and the SDSS $r-$band Petrosian magnitude from the SDSS DR12, $\mrpetro$, as a function of the stellar slope parameter $\slope$, of the galaxies in the SDSS sample.
Galaxies with a shallow slope (i.e. a low value of $\slope$) tend to have $\mrten > \mrpetro$: these are extended galaxies, for which their Petrosian radius (which is intended to capture most of the light of a galaxy) is larger than $10$~kpc, therefore the light enclosed within $10$~kpc is only a fraction of the total.
On the opposite end, the distribution in $\mrten-\mrpetro$ reaches negative values for very steep slopes, which is a result of the Petrosian radius (within which the Petrosian magnitude is defined) being smaller than $10$~kpc.
Most importantly, we can see that $\mrten-\mrpetro > -0.25$ for the vast majority of data points, with the exception of a few outliers (presumably catastrophic failures in photometric measurements in either SDSS or this work).
The $\mrpetro<19.4$ completeness limit of GAMA DR2 can then be translated into a corresponding limit of $\mrten<19.15$.
This is a conservative estimate, because for most galaxies the difference between $\mrten$ and $\mrpetro$ is larger than $-0.25$.

%One can then conclude that the SDSS main spectroscopic sample is nearly complete up to $\mrten < 17.45$.

%The SDSS main spectroscopic sample is more than 99\% complete up to an apparent $r-$band Petrosian magnitude $\mrpetro < 17.7$ \citep[][see their subsection 3.1 for definitions of Petrosian quantities]{Str++02}.
%SDSS Petrosian magnitudes, however, map nontrivially to HSC magnitudes in the context of this work. \Fref{fig:petro} shows the difference between the HSC $r-$band magnitude calculated within an aperture of $10$~kpc, $\mrten$, and the SDSS $r-$band Petrosian magnitude from DR12, $\mrpetro$, as a function of the stellar slope parameter $\slope$, of all galaxies in the sample. Galaxies with a shallower slope (i.e. a lower value of $\slope$) tend to have $\mrten > \mrpetro$: these are extended galaxies, for which their Petrosian radius (which is intended to capture most of the light of a galaxy) is larger than $10$~kpc, therefore the light enclosed within $10$~kpc is only a fraction of the total.
%On the opposite end, the distribution in $\mrten-\mrpetro$ reaches negative values for very steep slopes, which is a result of the Petrosian radius (within which the Petrosian magnitude is defined) being smaller than $10$~kpc.
%Most importantly, we can see that $\mrten-\mrpetro > -0.25$ for the vast majority of data points, with the exception of a few outliers (presumably catastrophic failures in photometric measurements in either SDSS or this work).
%One can then conclude that the SDSS main spectroscopic sample is nearly complete up to $\mrten < 17.45$.
%

%The corresponding completeness limit on $\mtenobs$ can be derived by considering the highest redshift 
Finally, I derived the corresponding completeness limit on $\mtenobs$ by generating, at the upper bound in the redshift distribution of the sample $z=0.20$, posterior predicted samples of $r-$band mass-to-light ratios from the same CSP models used for the stellar mass analysis of subsection \ref{ssec:mstar}, and then computing the value $\mtenobsmin$ for which 95\% of the samples are brighter than $\mrten = 19.15$.
The underlying assumption in this procedure is that the distribution in stellar population parameters does not vary significantly across the sample.
Using this procedure, I derived a value of $\log{\mtenobsmin}=10.79$: 95\% of elliptical galaxies of this mass at $z=0.20$, and an increasing fraction at lower redshifts, have an aperture $r-$band magnitude within $10$~kpc smaller than $19.15$, and are therefore included in the GAMA sample. Cutting the sample at $\mtenobs > \mtenobsmin$ then produces a nearly complete sample in $\mten$.

\end{document}